\documentclass{aa}

\usepackage{natbib}
\usepackage{graphicx}

\newcommand {\BD} {BD}
\newcommand {\BDs} {BDs}

\newcommand {\AV} {A_V}
 \newcommand {\chid} {\chi^2}

\newcommand {\VLM} {VLM}

\newcommand {\Ha} {H\alpha}

\newcommand {\Msun} {M_{\odot}}

\newcommand {\rss}{\mathcal{R}_{ss}}
\begin{document}

\title{17 new very low-mass members in Taurus}
\subtitle{The brown dwarf deficit revisited}

   \author{S.~Guieu
          \inst{1}
          \and
          C.~Dougados\inst{1}
          \and 
          J.-L.~Monin\inst{1,2}
          \and
          E.~Magnier\inst{3,4}
          \and
          E. L. Mart\'\i n\inst{5,6}
          }

   \offprints{Sylvain.Guieu@obs.ujf-grenoble.fr}

   \institute{Laboratoire d'Astrophysique de Grenoble, BP53, 38041 Grenoble, France.
               \and 
               Institut Universitaire de France
         \and
             Canada-France-Hawaii Telescope Corporation, P.O Box 1597,
             Kamuela, USA
         \and 
             University of Hawaii, Institute of Astronomy, 2680 Woodlawn Dr., Honolulu, HI  96821, USA
         \and 
          Instituto de Astrofisica de Canarias, E-38200 La Laguna, Tenerife, Spain; ege@iac.es. 
        \and
          University of Central Florida, Department of Physics, PO Box 162385, Orlando, FL 32816-2385, USA
        }

\abstract{
  Recent studies of the substellar population in the Taurus cloud have
  revealed a deficit of brown dwarfs compared to the Trapezium cluster
  population
  \citep{Briceno-1998,Luhman-2000,Luhman-2003a,Luhman-2004}. However, these
  works have concentrated on the highest stellar density regions of the
  Taurus cloud. We have performed a large scale optical survey of this
  region, covering a total area of  $\simeq 28\,{\rm deg}^2$, and
  encompassing the densest parts of the cloud as well as their surroundings,
  down to a mass detection limits of 15 M$_{\rm J}$. In this paper, we
  present the optical spectroscopic follow-up observations of  97 
  photometrically selected  potential new low-mass Taurus members,
   of which 27 are strong late-M spectral type ($SpT~\ge~M4V$) candidates. Our
  spectroscopic survey is  87 \% complete down to  $i^\prime=$20 for spectral types
  later than M4V, which corresponds to a mass completeness limit of 30
  M$_J$ for ages $\leq$ 10 Myr and Av $\le$ 4. We derive spectral types,
  visual absorption and luminosity class estimates and discuss our criteria
  to assess Taurus membership.  These observations reveal 5 new \VLM\
  Taurus members and 12 new \BDs. Two of the new \VLM\ sources and four of
  the new substellar members exhibit accretion/outflow signatures similar
  to the higher mass classical T Tauri stars. From levels of $\Ha$ emission
  we derive a fraction of accreting sources of 42 \% in the substellar
  Taurus population.  Combining our observations with previously published
  results, we derive an updated substellar to stellar ratio in Taurus of
  $\rss =0.23 \pm 0.05$. This ratio now appears consistent with the value
  previously derived in the Trapezium cluster under similar assumptions of
  $0.26 \pm 0.04$. %
  We find indication that the relative numbers of
  \BDs\ with respect to stars is decreased by a factor 2 in the central
  regions of the aggregates with respect to the more distributed
  population. Our findings are best explained in the context of the {\sl
  embryo-ejection model} where brown dwarfs originate from dynamical
  interactions in small N unstable multiple systems.
}

\authorrunning{S. Guieu et al.}
\titlerunning{17 new very low-mass members in Taurus}
\maketitle
\section{Introduction}

 Recent works by %
 \cite{Briceno-1998,Luhman-2000,Briceno-2002,Luhman-2003a,Luhman-2004},
 have revealed a factor of  1.4 to 1.8 deficit of Brown Dwarfs (\BDs)
 in the Taurus cloud compared to the Trapezium cluster, possibly indicating
 that sub-stellar object formation depends on the environment. If
 confirmed, the lower abundance of \BDs\ in Taurus could have strong
 implications on substellar Initial Mass Function (IMF) models.  However,
 as these previous studies were concentrated on high stellar density
 regions, the majority of the volume occupied by the molecular clouds was
 left unexplored.  There is a possibility that Taurus \BDs\ are not
 clustered in the same regions as most of the T Tauri stars, or that they
 have scattered away from their birth sites as proposed by
 \cite{Reipurth-2001}.  If \BDs\ are ejected during their formation, a
 significant fraction of the substellar content of the central parts of the
 cloud may have been missed. Indeed, with ejection velocities  $\ge$
 1\,km.s$^{-1}$, as predicted by some of the ejection models
 \citep{Reipurth-2001,Kroupa-2003a}, BDs could have travelled as far as 1$^{o}$ from
 their birth sites at the Taurus distance in $2-3\,$Myr.

   The recent availability of large visible and infrared cameras makes
   it possible to quickly survey large portions of the sky
   encompassing tens of square degrees down to the sensitivity needed
   to study the substellar domain. In order to check for the
   existence of a widespread \BD\ population in Taurus, we have used
   the Canada-France-Hawaii telescope to perform a large optical
   survey covering a total of $28\,{\rm deg}^2$ down to completeness
   limits of $i^\prime$=21.8, z$^{\prime}$=20.9, corresponding to a mass
   detection limit of 15 M$_{\rm J}$ for an age $\le$ 5 Myr and $\AV
   \le 5$, according to the pre-main sequence DUSTY model of
   \cite{Chabrier-2000}.  In this paper, we report on medium
   resolution optical spectroscopic follow-up observations of 97 
   photometrically selected Taurus candidate members, of which
   27 are strong late-M ($SpT~\ge~M4V$) candidates. This
   work is the continuation and completion of the preliminary spectroscopic study
   published in \cite{Martin-2001}, where initial candidates were selected
   from a $3.6\,{\rm deg}^2$ photometric survey conducted
   with the CFHT12k camera, which is a subset of the full $28\,{\rm
   deg}^2$ CFHT Taurus survey described here.

    In section~2, we first briefly describe the CFHT photometric survey and
   candidate selection procedures, then the spectroscopic
   observations presented here. We derive spectral types, reddenings and
   luminosity class estimates from both a spectral fitting procedure and
   the analysis of colors and Na {\sc i} equivalent widths in section~3. We
   assess Taurus membership in section~4 and discuss the properties of the
   new 5 \VLM\ members and 12 \BDs\ identified in this study. In
   particular, we investigate their accretion/outflow signatures and
   spatial distribution and update the stellar to brown dwarf ratio in
   Taurus. Finally, we discuss the implications of our results on the
   models for substellar object formation in section~5 and summarize our
   findings in section~6.

\section{Observations}

\subsection{The CFHT optical photometric survey \label{sec:photo}}

\begin{table*}[t]
\centering
\begin{tabular}{lccccllc}\hline
Instrument & FOV & Pixel & Date & Area            &      Band & Detection & Mass limit \\
  & (deg$^2$) & ($^{\prime\prime}$/pixel) & &  (deg$^2$)  &   & limit  & ($\AV=5$, 5Myr)    \\\hline\hline
CFHT12k & 0.33 & 0.21 & 1999-2001 & 3.6 & R, I, Z' & 23, 22, 21 & 15$M_{\rm J}$ \\
CFHT12k & 0.33 & 0.21 & 2002 & 8.8 & I, Z' & 22, 21 & 15$M_{\rm J}$ \\
Megacam & 1 & 0.19 & 2003-2004 & 24 & i', z' & 24, 23 & 10$M_{\rm J}$ \\\hline
\end{tabular}
\caption{Overview of the CFHT optical survey of the Taurus cloud}
\label{tab:phot-survey}
\end{table*}

\begin{figure*}[]
  \includegraphics[width=0.9\hsize]{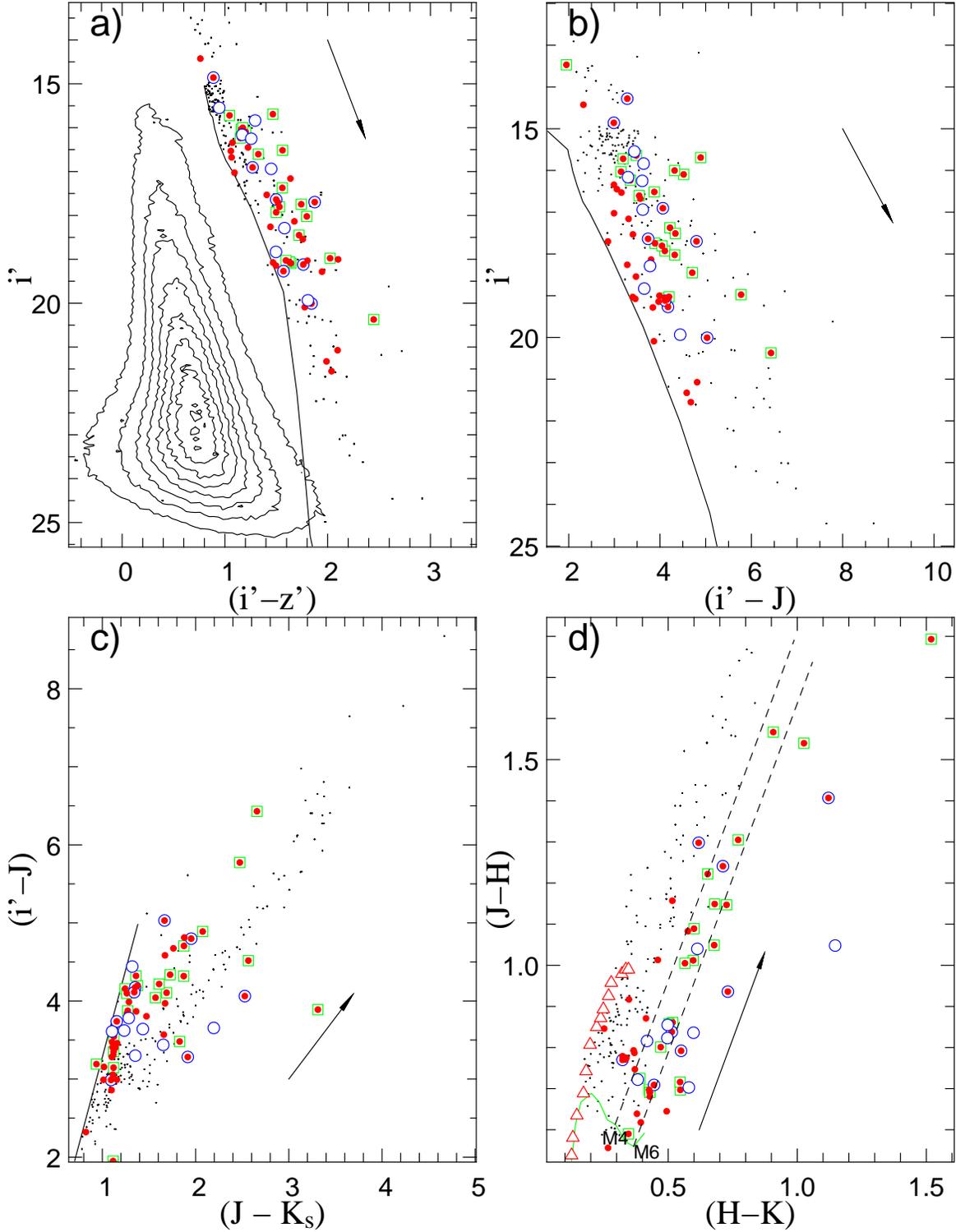}
  \caption{Observed  (not dereddened) 
  color-magnitude and color-color diagrams used to
  select low-mass Taurus candidates. %
  Small black dots in panels a), b)
  and c) are the 264 sources that pass through the selection process
  (see text for more details). Only the 227 sources with
  $i^\prime~<~20.$ are represented in panel d). The 44 (40 width
  $i^\prime<20$) primary photometric candidates observed
  spectroscopically are displayed in filled red circles. Open blue
  circles are previously known Taurus members. Green squares identify
  the spectroscopically confirmed new Taurus members from this work
  and \cite{Martin-2001}.  In panels a) and b), the full black curve
  shows the location of the 10 Myr isochrone from the DUSTY model of
  \cite{Chabrier-2000} at the Taurus distance. In panel c), the black
  curve shows the straight line approximating the dwarf sequence used
  to select M-type candidates.  The arrow indicates a reddening vector
  of $\AV$ = 4 magnitudes.  In the $(J-H)/(H-K)$ diagram are also
  shown the locations of the dwarf (green curve) and giant (open red
  triangles) branches, as well as reddening vectors extending from the
  M4 and M6 dwarf colors (dashed lines).
}
  \label{fig:sel}
\end{figure*}

The overall Taurus survey has been performed in 4 successive periods on the
Canada-France-Hawaii telescope with the CFH12k and MEGACAM large-scale
optical cameras (see Table~\ref{tab:phot-survey} for a detailed
journal). The technical characteristics of the CFH12k and MEGACAM cameras
are presented in \cite{Cuillandre-2000} and \cite{Boulade-2003}
respectively. The first part of the CFH12k survey (centered on the densest
part of the Taurus cloud, from 1999 to 2001), has been obtained as part of
CFHT director's discretionary time, while the remaining larger set of data
has been obtained, in service mode, as part of a larger key program devoted
to the study of young clusters. A detailed description of the optical
photometric survey is deferred to a forthcoming publication (Guieu et al
2005 in preparation). We summarize here its main characteristics.
 
Data reduction, performed at CFHT using elements of the Elixir
system \citep{Magnier-2004}, included bias and dark
subtraction, flat-fielding, fringing correction, bad pixels removal and
individual frame combination. Point source detection was performed on the
combined $I+z^{\prime}$ images. For the CFH12k data, PSF fitting photometry
was extracted with the PSFex routine from the SExtractor program
\citep{Bertin-1996}, while aperture photometry was obtained for the MEGACAM
data with the same program. Photometric catalogs were
combined, using the following transformation between CFH12k
($I$,$Z^{\prime}$) and MEGACAM ($i^{\prime}$,$z^{\prime}$) photometric
systems, computed from overlapping fields:

$$ i^{\prime} = I + 0.85 * (I-Z^\prime) + 0.23  $$ and $$
z^{\prime} = 0.25*I  + 0.746*Z^\prime + 0.32 $$

The total survey yielded more than $10^6$ sources detected down to $i^\prime$=24
and $z^\prime$=23. From the turn-over at the faint end of the magnitude
distribution, we estimate the completeness limits of our optical
photometric survey to be $i^\prime$=21.8 and $z^\prime$=20.9.
 Concerning the brightest sources in the survey, the saturation limits are
$i^\prime = 12.5$ and $z^\prime= 12$ 

 We identify Taurus low-mass candidate members with the following
procedure. We first select optical sources with a detection in the
near-infrared 2MASS catalogue at both J, H and K bands with quality flags A, B
or C. We then evaluate a rough estimate of the reddening vector towards
each source by assuming a maximum intrinsic $(J-H)_0$ color of 0.8, which
corresponds to a spectral type M9-L0 for a main sequence dwarf
\citep{Kirkpatrick-2000,Dahn-2002}. Reddening values estimated this way 
will be conservative lower estimates for earlier spectral type sources. Candidate
Taurus members are then selected according to their location in 
 the various  color-magnitude and color-color diagrams plotted in Figure \ref{fig:sel}. %
  We dereddened each candidate and required its position %
in the $i^\prime/(i^\prime-z')$ (or $I/(R-I)$ when available) and
$i^\prime/(i^\prime-J)$ color-magnitude diagrams to be compatible with stars younger than %
10~Myr, derived from comparison with the \cite{Chabrier-2000} pre-main
sequence tracks. We also require colors compatible with a dwarf M spectral type
in the  $(i^\prime-J)/(J-K)$ diagram.  
This selection procedure yielded 264 Taurus low-mass
candidates, of which 227 have $i^{\prime}$ band magnitudes $\leq$20 .

Strong galactic contamination from  primarily early M-type
background giants is expected.  In order to properly assess Taurus
membership, spectroscopic follow-up of the photometrically selected
candidates is therefore mandatory. We restrict ourselves to the
227 sources with $i^{\prime}$ band magnitudes $\leq$ 20, in order to
allow a proper spectroscopic follow-up in the optical domain. We
further use the $(J-H)/(H-K)$ diagram to separate early and late M
candidates. This procedure also allows to minimize the background
giant contamination, expected to be very strong at the early M
spectral types. Among the 227 sources with $i^{\prime}$ band
magnitudes $\leq$20, 47 sources appear later than M4V in the
$(J-H)/(H-K)$ (see Figure~\ref{fig:sel}d).

Among these 47 sources, are included 4 sources already published in
\cite{Martin-2001} (CFHT-Tau~1, CFHT-Tau~2, CFHT-Tau~3 and
CFHT-Tau~4) and 10 previously known Taurus low-mass members 
(V410 Tau Anon 13, ZZ Tau IRS, J04442713+2512164, J04381486+2611399, %
KPNO-Tau~9, %
KPNO-Tau~7, %
KPNO-TAU~5, %
KPNO-Tau~3, %
J04284263+2714039,
J04202555+2700355). We present in this paper medium
resolution (R$\simeq$ 1000) optical spectroscopic observations
obtained for 27 of the remaining 33 unknown late-M ($SpT~\ge~M4V$)
candidates including KPNO Tau 1-2-4 and 6 published in
\cite{Briceno-2002, Luhman-2003a}.  Thus, combining observations
presented in this work and previously published results, spectroscopic
information is available for 87 \% (41/47) of the Taurus candidates
later then M4V identified from our CFHT survey. We also included in
our spectroscopic survey 13 candidates with $SpT~<~M4V$, which
represent 7 \% (13/180) of the remaining early M-type candidates with
$i^\prime$ $<$ 20. A very strong galactic contamination is expected in
this latter sample.  We note that our spectroscopy survey is 100 \%
complete in the substellar domain: all the 13 sources that appear
later than M6V in the $(J-H)/(H-K)$ diagram are included in our
spectroscopic study.

 In addition to the 44 (40 with $i^\prime<20$) sources selected
with the procedure just described, we have also  observed in the
course of the different spectroscopic campaigns, 53 additional sources
selected with less restrictive criteria.
These include in particular %
sources selected from the $I/(R-I)$ diagram, observed at the
WHT in September 2000 but not published in \cite{Martin-2001}. All of
these additional objects, turn out to be spectroscopically confirmed
as non Taurus members (see \S \ref{sec:analysis}). They are
however included in this paper and  their derived properties are
summarized in Table~\ref{tab:params}.

Due to our selection procedure, the mass-completeness limit of our
spectroscopic sample is set both by the 2MASS completeness limits (J=15.25,
H=14.4, K=13.75) and the sensitivity limit of our optical spectroscopic
observations ($i^\prime\leq$20). 
Both are equally restrictive in terms of mass
completeness (see \S \ref{sec:rss}).

\subsection{Spectroscopy}

\begin{table*}[htb]
\centering
\begin{tabular}{llccllc}\hline
id & Instrument & Telescope & Dates & Wavelength & Dispersion  &  Number of sources\\
           &           &       & \AA        & \AA/pixel   &                   \\
\hline\hline
1 & FORS1  & VLT-UT3 & 2001-10-25 to 2002-02-20 & 6000-11000 & 2.59 & 10 \\
2 & FORS1  & VLT-UT1 & 2002-12-11 to 2003-03-05 & 6000-11000 & 2.59 & 10 \\
3 & LRIS   & Keck I & 2003-01-08, 2003-01-09 & 6000-10000 & 1.87 & 26 \\ 
4 & LRIS   & Keck I & 2004-12-03, 2004-12-04 & 6000-10000 & 1.87 & 32 \\ 
5 & ISIS   & WHT    & 2000-09-28, 2000-09-29 & 6490-9365 &  2.88 & 19 \\ \hline
\end{tabular}
\caption{Journal of spectroscopic observations}
\label{tab:spec-survey}
\end{table*}

Spectroscopic observations have been conducted with the  WHT/ISIS,
FORS1/VLT \citep{Appenzeller-1998} and LRIS/Keck instruments over 5
runs between September 2000 and December 2004. The detailed journal of the
spectroscopic observations is presented in Table~\ref{tab:spec-survey}.

\subsubsection{WHT/ISIS observations}

Spectroscopic observations were carried out using the
Intermediate-dispersion Spectrograph and Imaging System (ISIS) at the 4.2 m
William Herschel Telescope (WHT) in La Palma on 2000 September 2829. The
R158R grating on ISIS's red arm gave a wavelength coverage from 640.9 to
936.5 nm. The spectral resolution was 2.5 pixels (7.2 Å). The data were
reduced using standard routines for bias subtraction and flat-field
correction within the IRAF7 environment. Wavelength calibration was made
using the spectrum of an NeAr lamp. Instrumental response was calibrated
out using spectra of the flux standard Feige 24. A total of 29 sources 
were observed during this campaign, 4 were published in \cite{Martin-2001}
and 6 
turn out to be already known Taurus members. We publish here the 
remaining 19 sources.

\subsubsection{VLT/FORS observations}

FORS1\footnote{FOcal Reducer/low dispersion Spectrograph} observations were
performed in service mode between 2001 and 2003. In both cases, the grism 300I
was used, providing a dispersion of 2.59 \AA/pixel and a wavelength
coverage of 6000 to 11000 \AA. The slit width was 0.7$^{\prime\prime}$. The
standard ESO pipeline reduction procedure has been applied. This includes
bias correction, flat-fielding and wavelength calibration. Individual
spectra were extracted using the IRAF\footnote{IRAF distributed by National
Optical Observatories} {\it apextract} routine. The spectral response was
calibrated with observations of the standard M9 field \BD\ 
DENIS-P-J1048-3956 \citep{Delfosse-2001}.

\subsubsection{Keck/LRIS observations}

Two observing runs were conducted with the LRIS/Keck~I spectrograph in
January 2003 and December 2004. The spectral dispersion was 1.87~\AA/pixel
and wavelength coverage 6000-10000~\AA.  The slit width was
1.0$^{\prime\prime}$ for the first run and 1.5$^{\prime\prime}$ for the
second one, due to worse seeing conditions. Individual exposures were first
bias corrected and flat-fielded. Spectra were extracted with the IRAF {\it
apextract} routine, then wavelength calibrated with a reference arc lamp
exposure. The spectral response is calibrated with observations of the
following standards: LTT 377, Feige 110, BD+28 4211, BD+25 4655, PG
0918+029, G193-74, G191B2B, Feige 11.

\section{Data analysis \label{sec:analysis}}

\subsection{Spectral-type and reddening determination}
 
\subsubsection{Numerical spectral fitting \label{sec:fit}} 

\begin{figure*}[t]
  \includegraphics[width=\hsize]{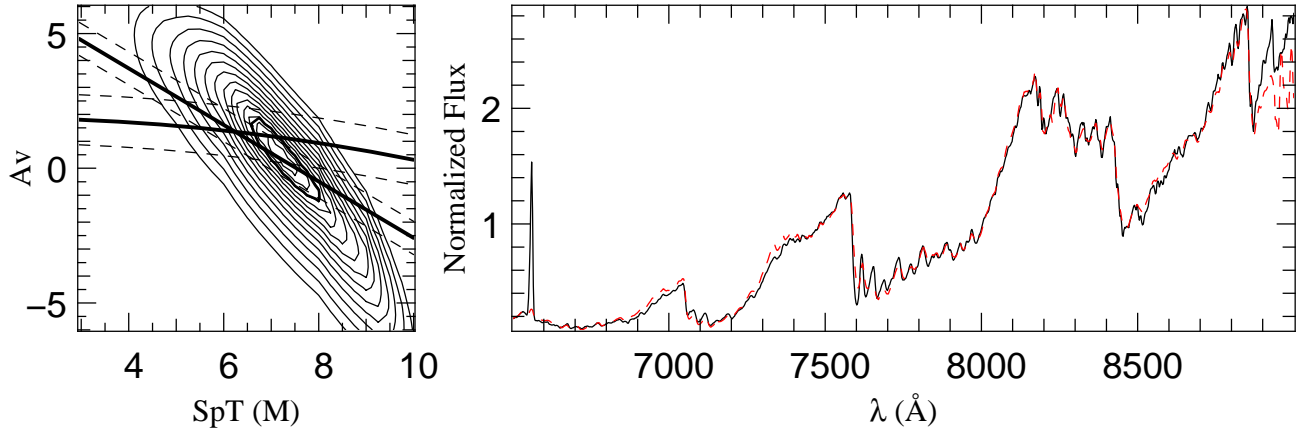}
  \caption{{\sl Left :} $\chi^2$ map obtained in
    the $\AV$-SpT plane for one of our candidate spectrum
    (2MASS~J04390396+2544264). In solid lines are also plotted the $\AV$
    variation {\it vs} SpT derived from the (I-J) and (J-H) colors using
    the \cite{Rieke-1985} extinction law. Dashed lines correspond to 1
    $\sigma$ uncertainties. The best fit obtained gives an M7.25 spectral type
    and $A_{V}=0.4\,$mag, compatible with the broad band colors. The best
    fit is obtained for the average of a dwarf and a giant template,
    indicating a luminosity class IV.
{\sl Right}: The resulting best fit ( average of a dwarf and giant template of SpT M7.25) is represented with a dashed red curve overlaid on the observed
    spectrum (solid black curve).}
  \label{fig:fit}
\end{figure*}

Spectral types can be derived from spectro-photometric indices, such as the ones
defined by \cite{Martin-1999}. However, these indices are affected by the
strong extinction expected towards Taurus members. From simulations of
artificially reddened standard spectra, we have computed that for a visual
extinction of 5 magnitudes, spectral types estimated from the PC3 indice
\citep{Martin-1999} are wrong by 2 classes for types around M7V.  We
therefore chose to apply a spectral fitting procedure similar to the one
presented in \cite{Luhman-1998}, allowing to derive simultaneously spectral
type and reddening.  We built a library of reference dwarf and giant
spectra spanning the range between M1 and L0.  The dwarfs reference library
was built from the combination of spectra published in
\cite{Kirkpatrick-1999}, \cite{Martin-1999} and \cite{Leggett-2000}, with 4
individual spectra, on average, for each M-spectral subclass.  The giant
reference library, built from a sample of 97 very bright M-giants, was
taken from \cite{Fluks-1994}. Half and quarter sub-classes were constructed
by combination of adjacent classes (for instance $M6.25 = < 0.75\times M6 +
0.25\times M7> $). For a given spectral type, each candidate spectrum can
be fitted either by the dwarf template, the giant template or the
average of both. This procedure therefore allows for an estimate of the
luminosity class. 
 
The fit is performed in the wavelength range 7000-8500~\AA, where the signal
to noise is maximized and the spectral response is best determined.
Figure~\ref{fig:fit} shows the $\chi^2$ map in the ($\AV$ - spectral type)
plane for one of our Taurus candidate (2MASS~J04390396+2544264)
with the corresponding best fit solution.  In the $\chi^2$ map, we have
also added the $\AV$ versus spectral type relations predicted by the (I-J)
and (J-H) colors. The calibration relations of broad band colors versus
spectral type are derived from linear fits to the photometry of M-type
dwarf standards observed by \cite{Leggett-1996,Leggett-1998,Leggett-2001}
and \cite{Dahn-2002}. In the example shown in Figure~\ref{fig:fit}, the
best fit solution found by our numerical fitting procedure appears
compatible, within the estimated 1$\sigma$ uncertainties, with the solution
predicted by the (I-J) and (J-H) colors. From simulations of artificially
reddened standard star spectra, we estimate typical internal uncertainties
for our numerical fitting procedure of half a class in spectral type and
0.8 magnitudes in $\AV$. Among our initial spectroscopic sample,  33 sources
are classified as giants, 38 as field dwarfs and 26 as sub-giants (ie best
represented by the average of a dwarf template and giant template). These
26 sub-giants appear as our primary Taurus pre-main sequence candidates. We
investigate below in more details the reliability of the parameters, in particular
reddenings and luminosity classes, determined by the spectral fitting
procedure.
 
\subsubsection{Reddening estimates from color-color diagrams \label{sec:reddening}}

\begin{figure*}[!t]
\begin{tabular}{cp{.05\hsize}c}
\includegraphics[width=0.4\hsize]{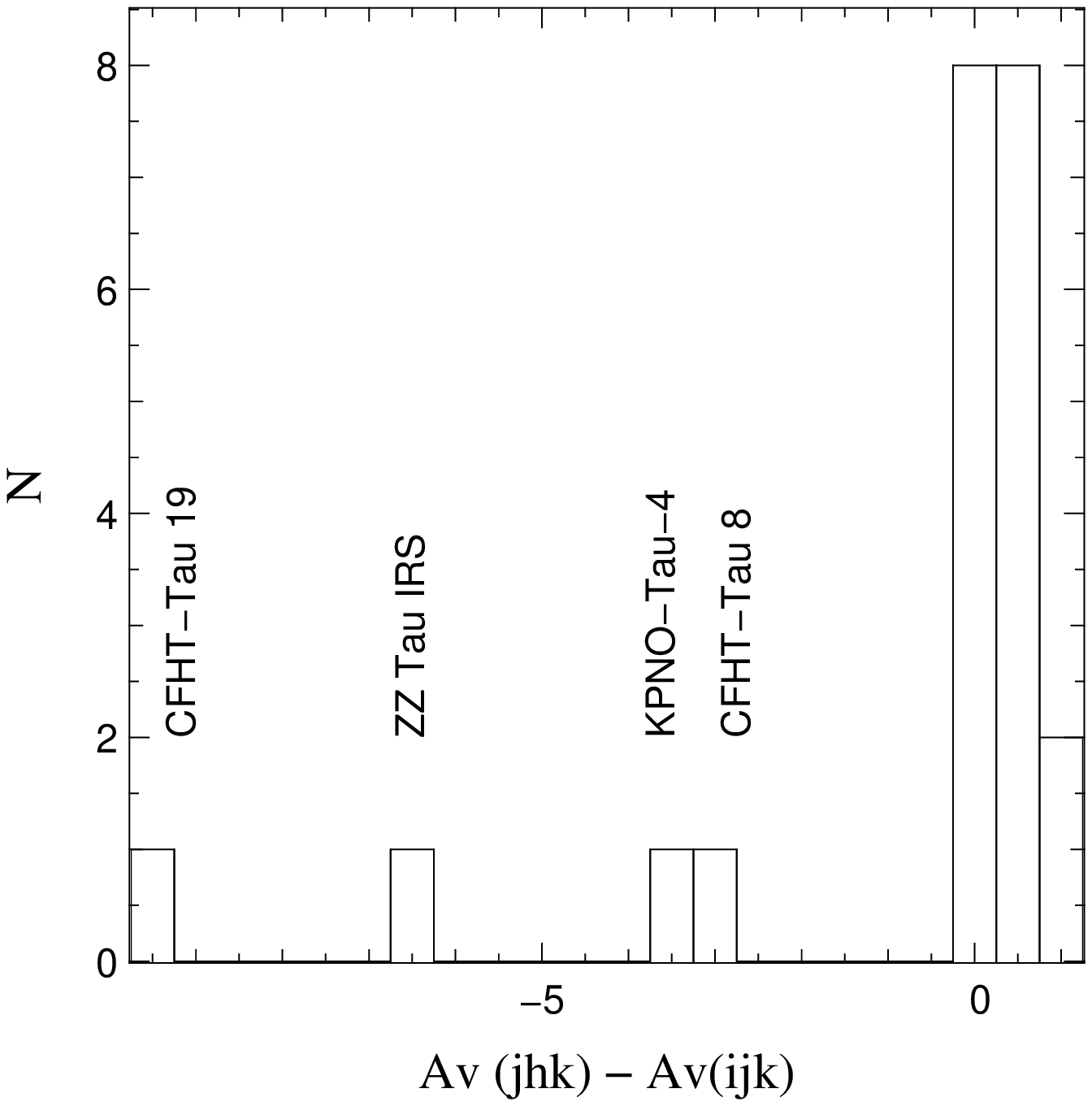} &&
\includegraphics[width=0.4\hsize]{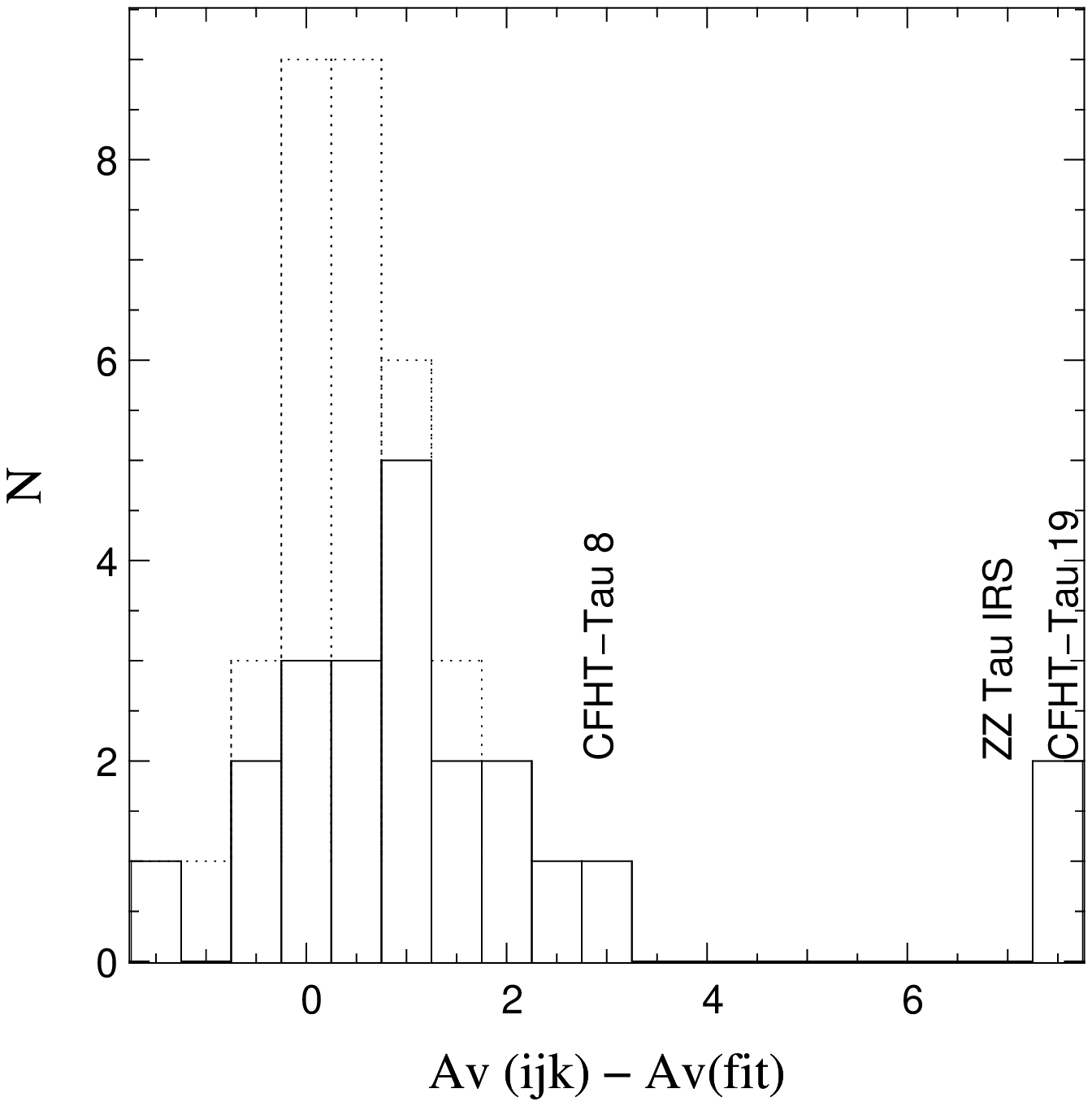}
\end{tabular}
\caption{Comparison of $\AV$ values derived from the color-color diagrams
  and the spectral fitting procedure. {\sl Left}: Histogram of differences
  in $\AV$ values derived from the (J-H)/(H-K) and from the (I-J)/(J-K)
  diagrams. Sources classified as giants have been excluded.  {\sl Right}:
  Histograms of differences in $\AV$ values derived from the (I-J)/(J-K)
  colors and from the spectral fitting procedure, in {\sl full line} for
  sources classified as sub-giants, in {\sl dotted line} for sources classified
  as field dwarfs.  Individual sources with the largest $\AV$ discrepancies
  are labelled and discussed in the text.}
\label{fig:histav}
\end{figure*}

In order to assess in more details the reliability of the $\AV$ values
derived from our numerical fitting procedure, we independently
estimate extinction values from the (J-H)/(H-K) and (I-J)/(J-K)
color-color diagrams.  We exclude from the following analysis the
33 sources classified as giants by the spectral fitting
procedure. Most of these turn out to be early type (SpT $\le$ M2III)
background giants.  We are therefore left with a sample of 64
sources classified either as M dwarfs or sub-giants. Sources with no
detectable excess infrared emission in the (J-H) versus (H-K) diagram
are dereddened back to the locus of dwarfs, while those lying to the
right of the reddened dwarf sequence in the near-IR color-color
diagram are dereddened back to the locus of actively accreting T Tauri
stars, as defined by \cite{Meyer-1997}. In the (I-J) versus (J-K)
diagram, sources are dereddened back to the dwarf sequence.

We compare in Figure~\ref{fig:histav} ({\sl Left}) $\AV$ values derived in
these two ways. Four sources (individually labelled) show large differences
in excess of 2 magnitudes. Two of these sources also show significant
near-infrared excess emission in the (J-H)/(H-K) diagram
(Figure~\ref{fig:jhhk}), indicating that the determination of $\AV$ is
likely very uncertain.  Except for these 4 sources, $\AV$ values derived
from both color-color diagrams agree remarkably well: the distribution of
differences peaks at $\Delta \AV$=0 with a dispersion of 0.3 magnitudes. On
the other hand, we find a systematic difference between the $\AV$ values
obtained from the spectral fitting procedure and the ones derived from the
color-color diagrams (Figure~ \ref{fig:histav}, {\sl Right}), especially
for the sources classified as sub-giants: the distribution of differences
peaks at $\Delta \AV$=1 with a dispersion of 0.3 magnitudes. For sources
classified as dwarfs the distribution of differences peaks at $\Delta
\AV$=0.25 with a dispersion of 0.3 magnitudes. 

 The discrepancy observed in $\AV$ determinations for the sub-giant
population likely arises from an improper calibration of the intrinsic
broad-band colors for sources with surface gravities intermediate between
dwarfs and giants. Indeed the derivation of reddening values from
main-sequence colors assumes that a young object has the same intrinsic
continuum slope as a dwarf. However, \cite{Luhman-1999} noted that
intrinsic $(I-J)$ colors of pre-main sequence sources are significantly
redder than the ones of dwarfs at spectral types later than M5V. Indeed,
the spectra of the mid to late M young stellar objects always rise more
rapidly beyond 8000\AA\ than the spectra of main-sequence dwarfs of the
same spectral type.  The derivation of reddening values from dwarf colors
involving the I band will therefore systematically overestimate $\AV$ for
young mid to late-M stars. Our analysis seems to indicate that this is also
true for reddening values derived from near-infrared colors. %
The spectral fitting procedure on the other hand takes into account
the fact that the surface gravity of young stars is intermediate
between the ones of dwarfs and giants. Although we note that the
continuum slope of a young star may not be exactly the average of a
dwarf and giant continuum, we estimate that the Av values derived from
the spectral fitting procedure are less biased and we adopt them in
the following.

\subsection{Luminosity class and Sodium equivalent widths \label{sec:class}}

\begin{figure}[htb]
   \includegraphics[width=\hsize]{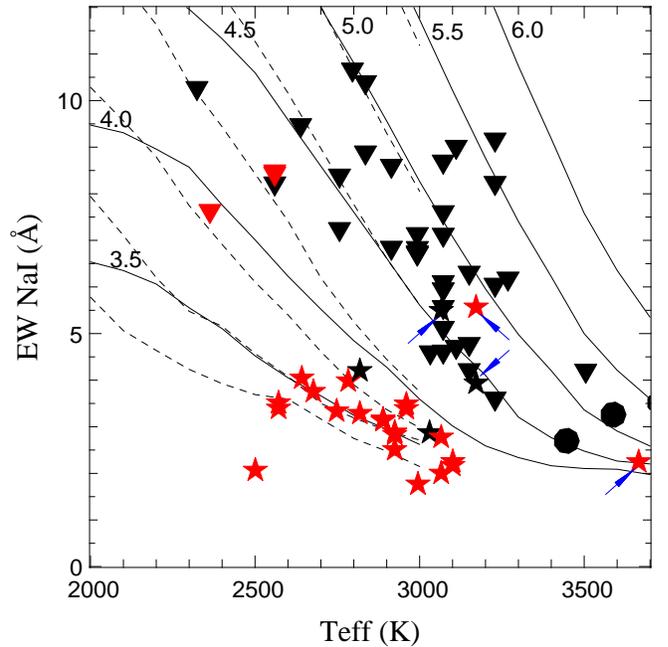} \caption{Measured EW(Na
   {\sc i}) {\it vs} effective temperature for our spectroscopic
   sample. {\it Star} symbols represent spectra best fitted
   with the average of a giant and field dwarf template (sources
   classified as luminosity class IV). {\it triangles} show
   sources best fitted with a field dwarf template; {\it circles}
   correspond to a giant template.  Sources with EW(H$\alpha$) $\ge$
   8~ \AA\ are plotted in red.  {\it Solid} lines show predictions
   from the DUSTY models from \cite{Chabrier-2000} spanning
   $\log_{10}({\rm g}/{\rm g}_0) = $3.5 to 6 by step of 0.5 (as
   labelled); {\it dashed} lines correspond to the COND model spanning
   $\log_{10}({\rm g}/{\rm g}_0) = $2.5 to 6. The 4 sources with
   discrepant EW(Na {\sc i}) are indicated with an arrow.}
\label{fig:na}
\end{figure}

The Na{\sc i} doublet at 8183 and 8195 \AA~ is sensitive to surface gravity
effects \citep{Martin-1999} for effective temperatures in the range 2500
to 3500~K. We therefore compare the luminosity class estimated for
our sources from the spectral fitting procedure with Na {\sc i} equivalent
widths measurements. The EW(Na{\sc i}) is computed between 8172~\AA\ and
8207~\AA\ (the doublet is not fully resolved at the moderate spectral
resolution of our observations). The local continuum is estimated by a
linear fit to the three highest points in each of the following spectral
intervals: [8090:8172 \AA] and [8207:8280 \AA]. We estimate typical
uncertainties on the equivalent widths to be on the order of 5 \%.

Figure~\ref{fig:na} illustrates the excellent agreement between the
luminosity class estimated from our spectral fitting procedure and the
Na{\sc i} equivalent widths.  In this figure, we plot the measured
EW(Na{\sc i}) versus effective temperature T$_{\rm eff}$. Spectral types,
derived from the spectral fitting procedure, are converted into effective
temperatures using the relationships given by \cite{Luhman-1999} for
pre-main sequence stars, dwarfs and giants respectively. Eighteen sources
classified as early-type ($SpT <$ M1) giants are not included in the graph.
Figure ~\ref{fig:na} shows a clear separation between sources
spectroscopically classified as field dwarfs and sources identified as
sub-giants. All the sources classified as field dwarfs show EW(Na{\sc
i}) comparable to the predictions of the model atmospheres from
\cite{Chabrier-2000} with surface gravities $\log_{10}({\rm g}/{\rm g}_0) \geq $ 4. On
the other hand, all but  4 of the estimated luminosity class IV sources show
EW(Na{\sc i}) consistent with predictions from the model atmospheres for
low surface gravities ($\log_{10}({\rm g}/{\rm g}_0) \leq $ 4). In the following, we
therefore adopt the luminosity class estimated from the spectral fitting
procedure and discuss the nature of these  4 discrepant low surface gravity
candidates (indicated by an arrow in Figure~\ref{fig:na}). We notice that
one known Taurus \BD\ (KPNO-Tau~4) shows an unusually small Na{\sc i}
equivalent width (2.07 \AA) for its estimated effective temperature ($\leq$
2500~K). However, the detailed profile of the Na{\sc i} doublet in this
source appears to differ significantly from both the model predictions and
the corresponding dwarf template line profile, suggesting a core emission
which could explain its low position in Figure~\ref{fig:na}.
 
\begin{table*}
  
  \centering 
   {\scriptsize

\begin{tabular}{ lclllllll } \hline\hline
 2MASS Name&Run&RA(J2000)&DEC(J2000)&Av&EW(Na)&EW($H_\alpha$)&SpT& 
 \\ \hline
\multicolumn{9}{c}{{\normalsize Field stars}} \\ \hline {J04015740+2626199$*$}&4&04:01:57.409&26:26:19.993&1.59&4.20&0.55&2.00&V\\
{J04130888+2805559$*$}&1&04:13:08.887&28:05:55.943&0.00&8.39&1.69&6.75&V\\
{J04141928+2804118$*$}&5&04:14:19.283&28:04:11.813&0.47&6.18&0.58&3.50&V\\
{J04141951+2757014$*$}&1&04:14:19.519&27:57:01.451&0.90&8.60&-7.90&5.75&V\\
{J04170178+2821593$*$}&5&04:17:01.787&28:21:59.375&8.39&3.62&0.00&1.00&III\\
{J04173746+2811230$*$}&5&04:17:37.463&28:11:23.046&9.82&2.77&0.00&1.00&III\\
{J04180215+2817487$*$}&1&04:18:02.157&28:17:48.790&4.74&8.69&-1.10&4.75&V\\
{J04180537+2828011$*$}&5&04:18:05.374&28:28:01.114&8.00&1.67&0.00&1.00&III\\
J04202471+2703532&4&04:20:24.713&27:03:53.219&6.14&1.39&3.94&1.00&III\\
J04245748+2402016&4&04:24:57.485&24:02:01.687&0.46&9.47&-5.26&7.50&V\\
J04274047+2610193&2&04:27:40.473&26:10:19.340&4.18&3.68&0.91&1.00&III\\
{J04274721+2602405$*$}&2&04:27:47.210&26:02:40.531&1.23&10.27&-4.25&9.50&V\\
{J04280147+2618101$*$}&3&04:28:01.440&26:18:10.440&2.00&1.02&2.90&1.00&III\\
J04281440+2622409&3&04:28:14.400&26:22:40.800&2.36&1.41&2.52&1.00&III\\
{J04281884+2623169$*$}&3&04:28:18.840&26:23:16.800&2.41&1.15&1.36&1.00&III\\
{J04282842+2608177$*$}&3&04:28:28.425&26:08:17.700&2.65&3.26&0.67&3.00&III\\
J04284464+2424202&4&04:28:44.644&24:24:20.214&0.00&8.89&-7.19&6.25&V\\
{J04290207+2558061$*$}&3&04:29:02.074&25:58:06.132&0.10&4.71&-0.36&4.50&V\\
{J04291217+2618592$*$}&3&04:29:12.168&26:18:59.760&2.66&1.16&0.82&1.00&III\\
{J04294344+2436224$*$}&3&04:29:43.446&24:36:22.475&4.18&9.01&0.81&4.50&V\\
{J04294651+2431493$*$}&5&04:29:46.519&24:31:49.328&10.48&1.48&0.00&1.00&III\\
J04294723+2529188&4&04:29:47.236&25:29:18.892&0.00&8.43&-23.70&8.00&V\\
J04295622+2503559&4&04:29:56.228&25:03:55.904&2.25&4.63&-3.56&1.00&V\\
{J04300120+2548459$*$}&3&04:30:01.207&25:48:45.914&0.09&3.60&1.49&3.75&V\\
J04301981+2630196&4&04:30:19.810&26:30:19.667&1.56&6.10&-5.87&4.75&V\\
{J04304498+2551071$*$}&3&04:30:45.000&25:51:07.920&1.62&1.63&1.06&1.00&III\\
{J04312215+2449527$*$}&4&04:31:22.150&24:49:52.730&1.56&5.12&0.77&4.75&V\\
{J04312630+2544327$*$}&3&04:31:26.304&25:44:32.640&0.85&1.94&0.78&1.00&III\\
{J04312740+2553281$*$}&3&04:31:27.408&25:53:28.320&2.22&0.97&1.41&1.00&III\\
{J04312868+2417184$*$}&5&04:31:28.689&24:17:18.442&5.34&5.60&0.00&1.00&III\\
{J04313322+2419015$*$}&5&04:31:33.229&24:19:01.556&3.70&8.23&0.00&3.75&V\\
J04313986+2436357&3&04:31:39.864&24:36:36.360&4.44&1.68&1.43&1.00&III\\
{J04314390+2409418$*$}&5&04:31:43.901&24:09:41.857&5.42&4.15&2.09&1.00&III\\
J04315468+2436153&3&04:31:54.672&24:36:16.200&7.07&1.91&1.97&1.00&III\\
J04315728+2436138&3&04:31:57.288&24:36:14.760&5.97&2.91&1.28&1.00&III\\
{J04315921+2423190$*$}&5&04:31:59.219&24:23:19.021&2.96&9.17&0.00&3.75&V\\
{J04320212+2412431$*$}&5&04:32:02.124&24:12:43.186&6.46&4.98&0.00&1.00&III\\
{J04321076+2420377$*$}&5&04:32:10.764&24:20:37.788&6.33&4.08&2.94&1.00&III\\
{J04321441+2456174$*$}&4&04:32:14.418&24:56:17.408&0.00&6.72&1.60&5.25&V\\
{J04322154+2411221$*$}&1&04:32:21.546&24:11:22.106&0.00&8.48&-16.50&8.00&V\\
{J04330191+1831066$*$}&3&04:33:01.917&18:31:06.654&3.01&3.52&0.87&2.00&III\\
{J04330277+1824258$*$}&3&04:33:02.832&18:24:26.280&0.34&1.06&1.70&1.00&III\\
J04334207+2217333&4&04:33:42.077&22:17:33.320&0.10&6.84&1.10&5.25&V\\
{J04334314+2647424$*$}&4&04:33:43.147&26:47:42.482&2.45&4.22&-0.52&4.25&V\\
{J04341429+2416062$*$}&1&04:34:14.298&24:16:06.240&2.07&4.60&-5.26&5.00&V\\
{J04343346+2253070$*$}&2&04:34:33.465&22:53:07.051&1.17&8.22&-4.04&8.00&V\\
{J04345366+2410538$*$}&5&04:34:53.665&24:10:53.803&0.02&10.40&-5.73&6.25&V\\
J04350520+2250385&3&04:35:05.184&22:50:39.120&3.41&1.39&0.40&1.00&III\\
{J04351784+2242282$*$}&1&04:35:17.844&22:42:28.238&1.35&7.61&-4.99&4.75&V\\
{J04351943+2238203$*$}&5&04:35:19.434&22:38:20.386&4.49&3.83&0.00&1.00&III\\
J04353690+2640182&4&04:35:36.909&26:40:18.250&5.58&3.19&0.84&1.25&III\\
{J04353912+2240369$*$}&5&04:35:39.129&22:40:36.934&1.13&6.31&-7.21&4.25&V\\
{J04355699+2713092$*$}&4&04:35:57.000&27:13:09.250&0.00&7.14&-3.24&5.25&V\\
{J04362765+2239496$*$}&1&04:36:27.653&22:39:49.651&0.21&10.67&-0.72&6.50&V\\
{J04371221+2439036$*$}&4&04:37:12.218&24:39:03.618&0.10&6.84&-3.06&5.75&V\\
J04381840+2455135&4&04:38:18.405&24:55:13.505&0.00&5.57&-4.36&4.75&V\\
J04384651+2558153&3&04:38:46.512&25:58:14.880&3.97&0.81&3.72&1.00&III\\
{J04392692+2552592$*$}&5&04:39:26.920&25:52:59.286&15.44&2.70&0.00&4.00&III\\
{J04393591+2542325$*$}&2&04:39:35.916&25:42:32.598&0.00&7.25&1.12&6.75&V\\
{J04395172+2549182$*$}&5&04:39:51.726&25:49:18.282&10.33&2.82&0.00&1.00&III\\
J04401040+2556029&3&04:40:10.368&25:56:03.120&3.75&7.12&-1.47&4.75&V\\
{J04402392+2554478$*$}&2&04:40:23.927&25:54:47.808&4.61&6.74&-5.05&5.25&V\\
{J04404085+2542054$*$}&3&04:40:40.824&25:42:05.400&3.18&0.99&3.51&1.00&III\\
J04411647+2507537&3&04:41:16.464&25:07:53.760&2.81&2.33&2.58&1.00&III\\
{J04411897+2515090$*$}&2&04:41:18.979&25:15:09.043&0.29&7.63&-13.08&9.25&V\\
{J04413108+2519222$*$}&5&04:41:31.085&25:19:22.300&10.93&3.26&0.00&1.00&III\\
J04422975+2519485&2&04:42:29.752&25:19:48.508&5.16&5.95&-7.44&4.75&V\\
J04425194+2414525&4&04:42:51.946&24:14:52.598&0.51&4.78&0.85&4.25&V\\
{J04430638+2523237$*$}&1&04:43:06.382&25:23:23.766&3.14&6.05&-3.39&3.75&V\\
{J04430807+2522103$*$}&3&04:43:08.040&25:22:10.560&3.72&2.88&0.38&1.00&III\\
{J04435916+2419389$*$}&4&04:43:59.169&24:19:38.989&1.09&4.62&-3.81&4.75&V\\
\hline \multicolumn{9}{c}{{\normalsize Uncertain membership}} \\ \hline J04272297+2636474&4&04:27:22.975&26:36:47.498&2.41&5.49&-6.94&5.50&IV\\
{J04315129+2506524$*$}&4&04:31:51.298&25:06:52.488&1.86&3.94&-2.62&4.75&IV\\
{J04355760+2253574$*$}&5&04:35:57.608&22:53:57.491&2.55&5.57&-21.58&4.75&IV\\
 \hline

\end{tabular}
 }

  \caption{Derived parameters and spectral
  properties of the sources identified as field stars or with
  uncertain Taurus membership. The first column gives the 2MASS name.
  The second column corresponds to the observation run as indicated in
  Table \ref{tab:spec-survey}. The following two columns are
  2MASS coordinates. Reddening values, spectral type and luminosity
  class derived from the spectral fitting procedure (see \S
  \ref{sec:fit}) are listed in columns (5), (8) and (9)
  respectively. Columns (6) and (7) give the measured Na {\sc i} and
  $\Ha$ equivalent widths.  The 2MASS names followed with a $*$
  correspond to candidates not selected with the primary selection
  method (see \S \ref{sec:photo}).} 
\label{tab:params}
\end{table*}

\begin{table*}
  \centering

   {\scriptsize

\begin{tabular}{ llclllllll } \hline\hline
 2MASS Name&Name&Run&RA(J2000)&DEC(J2000)&Av&EW(Na)&EW($H_\alpha$)&SpT& 
 \\ \hline
J04151471+2800096&KPNO-Tau-1&1&04:15:14.714&28:00:09.612&0.41&3.40&-26.82&9.00&IV\\
J04174955+2813318&KPNO-Tau 10&3&04:17:49.554&28:13:31.854&0.00&2.78&-40.09&5.50&IV\\
J04185115+2814332&KPNO-Tau-2&5&04:18:51.156&28:14:33.241&0.37&3.16&-8.39&6.75&IV\\
J04210795+2702204&CFHT-Tau 19&4&04:21:07.953&27:02:20.418&7.34&2.17&-442.17&5.25&IV\\
J04214631+2659296&CFHT-Tau 10&4&04:21:46.315&26:59:29.609&3.59&3.49&-17.32&6.25&IV\\
J04221644+2549118&CFHT-Tau 14&4&04:22:16.440&25:49:11.842&0.57&3.34&-48.77&7.75&IV\\
J04221675+2654570&CFHT-Tau 21&4&04:22:16.759&26:54:57.078&6.61&2.25&-45.09&1.25&IV\\
J04242646+2649503&CFHT-Tau 9&4&04:24:26.462&26:49:50.362&0.91&3.39&-9.95&6.25&IV\\
J04272799+2612052&KPNO-Tau-4&2&04:27:27.997&26:12:05.270&2.45&2.07&-158.08&9.50&IV\\
J04274538+2357243&CFHT-Tau 15&4&04:27:45.380&23:57:24.325&1.30&3.76&-18.90&8.25&IV\\
J04292165+2701259&CFHT-Tau 18&4&04:29:21.653&27:01:25.946&4.85&1.77&-38.30&6.00&IV\\
J04295950+2433078&CFHT-Tau 20&4&04:29:59.508&24:33:07.852&3.60&2.00&-114.03&5.50&IV\\
J04300724+2608207&KPNO-Tau-6&2&04:30:07.244&26:08:20.792&0.88&3.51&-207.91&9.00&IV\\
J04302365+2359129&CFHT-Tau 16&4&04:30:23.655&23:59:12.991&1.51&4.05&-16.76&8.50&IV\\
J04305171+2441475&ZZ Tau IRS&4&04:30:51.714&24:41:47.512&2.38&2.25&-141.13&5.25&IV\\
J04312669+2703188&CFHT-Tau 13&4&04:31:26.690&27:03:18.810&3.49&4.20&-3.60&7.25&IV\\
J04321786+2422149&CFHT-Tau 7&3&04:32:17.862&24:22:14.984&0.00&2.83&-8.63&6.50&IV\\
J04325026+2422115&CFHT-Tau 5&2&04:32:50.265&24:22:11.564&9.22&3.98&-29.84&7.50&IV\\
J04330945+2246487&CFHT-Tau 12&4&04:33:09.457&22:46:48.702&3.44&2.51&-79.65&6.50&IV\\
J04350850+2311398&CFHT-Tau 11&4&04:35:08.508&23:11:39.865&0.00&3.13&-45.07&6.75&IV\\
J04390396+2544264&CFHT-Tau 6&3&04:39:03.960&25:44:26.416&0.41&3.29&-63.74&7.25&IV\\
J04400174+2556292&CFHT-Tau 17&1&04:40:01.745&25:56:29.227&6.50&2.87&-7.26&5.75&IV\\
J04411078+2555116&CFHT-Tau 8&3&04:41:10.785&25:55:11.651&1.77&2.89&-52.00&6.50&IV\\
 \hline

\end{tabular}
 }
  \caption{Same as Table~\ref{tab:params} for confirmed Taurus members.}
  \label{tab:params2}
\end{table*}
\section{Results}

\begin{figure*}[]
\includegraphics[width=\hsize]{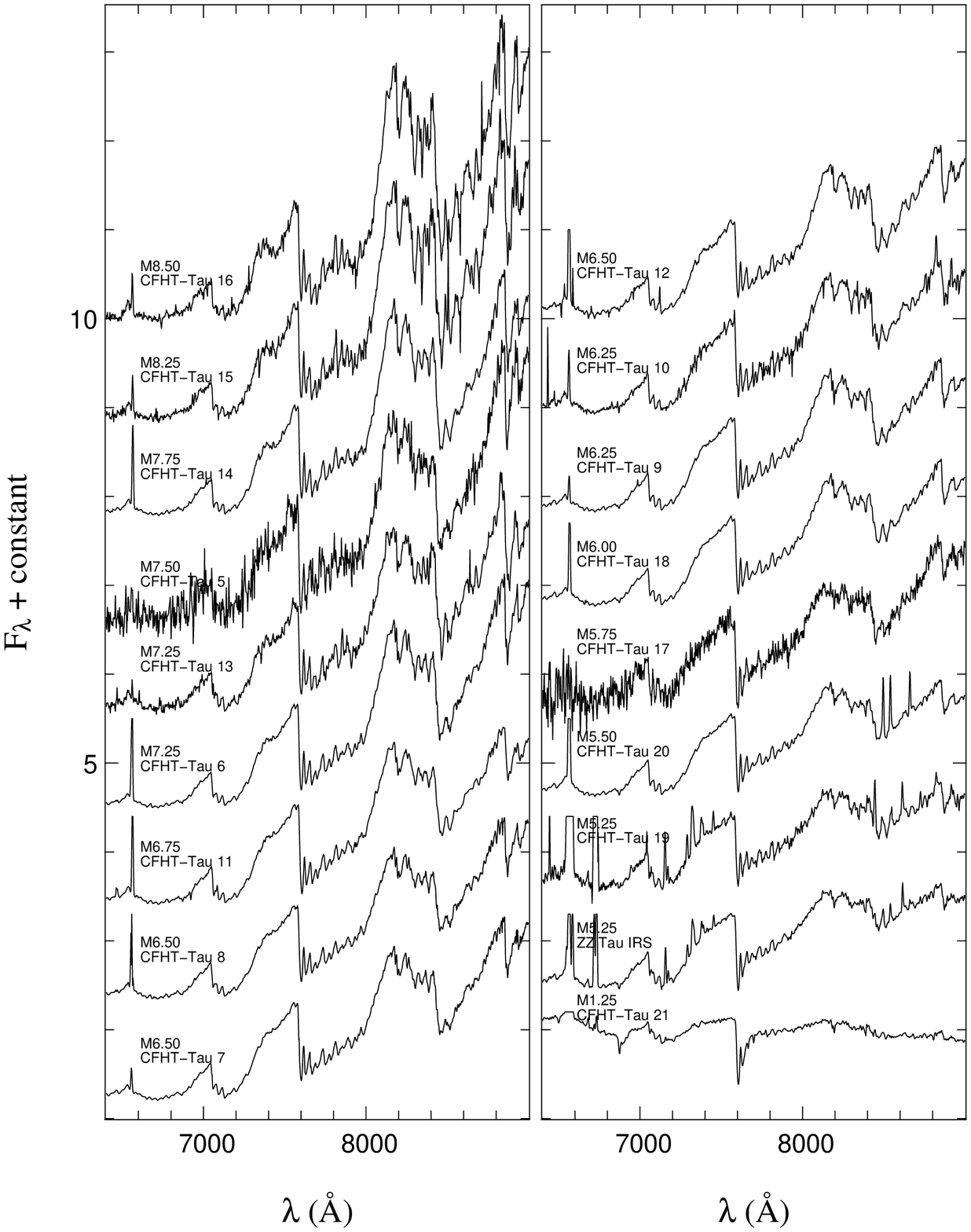}
\caption{Medium-resolution optical spectra of the new Taurus very low-mass members
  and brown dwarfs. The spectra have been corrected for extinction. The
  derived spectral type is indicated. The spectrum of ZZ~Tau~IRS is also shown.}
\label{fig:spec}
\end{figure*}

\subsection{Identification of Taurus members}

We compile in Table~\ref{tab:params}  and \ref{tab:params2}  the
reddening, spectral type and luminosity class derived from the
spectral fitting for our  97 spectroscopic sources. We also list
measured equivalent widths in Na{\sc i} and H$\alpha$.  We use the
luminosity class as the primary criterion to identify Taurus pre-main
sequence sources. 26 sources are classified as luminosity class IV
through our spectral fitting procedure.  Among these, only 4
show EW(Na {\sc i}) indicative of larger surface gravities
$\log_{10}({\rm g}/{\rm g}_0) >$ 4, as determined from
Figure~\ref{fig:na}.

The first one, 2MASS~J04221675+2654570, has an early spectral type
(M1.25), and falls in the temperature regime where the sodium
equivalent width is no longer sensitive to surface gravity for
moderate values ($\log_{10}({\rm g}/{\rm g}_0)\leq$ 5). In addition,
this source shows significant H$\alpha$ emission (with an equivalent
width of -24.5~\AA), reddening ($\AV$ = 6.6) and near-infrared excess
(see Fig. \ref{fig:jhhk}). We therefore classify it as a Taurus
member.

 The second is 2MASS~J04355760+2253574. The spectrum of this
object is best fitted by the average of a M4.75 dwarf and giant with a
reddening of $\AV$ = 2.55 and it displays $\Ha$ emission with an
equivalent width of -21.58 \AA, indicating a possible pre-main
sequence nature. Its near-infrared colors ((J-H)=1.01 and (H-K)= 0.56)
suggest significant reddening and/or near-infrared excess emission.
However, its measured EW(Na{\sc i}) of 5.25~\AA\ is more representative
of field dwarf surface gravities. In addition, when placed on the HR
diagram at the Taurus distance with the above derived parameters, it
falls close to the 100 Myr isochrone, significantly below the Taurus
population. The spectral fit obtained with a dwarf template gives a
spectral type of M5.25 and $\AV$ of 0.45 with a $\chid$ of
1.17 (see Figure~\ref{fig:na}). However a foreground nature does not seem compatible with the
near-infrared colors. It is therefore difficult at this stage to
conclude on the nature of this source.
 
\begin{figure*}
  \includegraphics[width=\hsize]{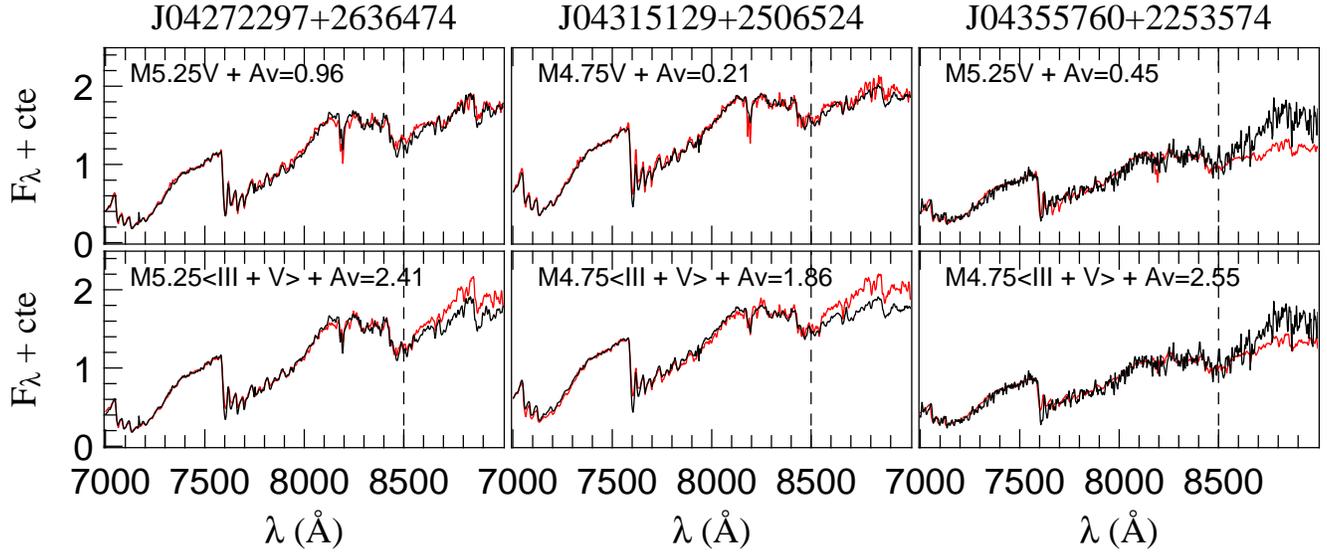} \caption{ Spectral
  fits (red thick lines) obtained for the 3 sources (black thin lines)
  with uncertain membership. The top panels represent the best
  solution obtained with a dwarf template while the bottom panels show
  the best fits obtained with the average of a dwarf and giant. The
  vertical dashed lines represent the wavelength limits used for the
  fit (7000-8500 \AA).}  
\label{fig:embigous}
\end{figure*}

The remaining two sources, 2MASS~J04272297+2636474 and
2MASS~J04315129+2506524, have mid-M spectral types, low-level line
emission (EW(H$\alpha$) emission $<$8~\AA) and occupy in
Figure~\ref{fig:na} the same location as dwarfs. For these two
sources, spectral types M5.5 and M4.75 and $\AV$ values of 2.41 and
1.86 are derived from the best fit solution obtained with the average
of a dwarf and giant template. Their derived location in the HR
diagram with the above parameters is shown in
Figure~\ref{fig:hr}. Both sources fall between the 10 and 30 Myr
isochrones, at the low luminosity end of the Taurus population.
However, as illustrated in Figure~\ref{fig:embigous}, the fits
obtained with a dwarf template alone appear in both cases very close
to the previous best fit solution.  The differences in $\chid$
between the dwarf and sub-giant fits for these two sources are 0.06
and 0.35 respectively, significantly lower than the average of the
$\chid$ differences for the remaining Taurus candidate population of
1.8$\pm$1.25. The dwarf fits give a similar spectral type but
significantly lower reddening values compatible, within our derived 1
$\sigma$ uncertainty of 0.8 mag, with a foreground field dwarf nature.
Their near-infrared colors ((J-H)=0.35, (H-K)=0.7-0.8) and I band
magnitudes appear also compatible with a foreground nature.

We therefore classify these last 3 sources in Table~\ref{tab:params} with uncertain  membership and do not consider them as Taurus members in the following. 
We are therefore left with 23 bonafide Taurus members. We note
that all but 2 of these show strong H$\alpha$ emission with equivalent
widths in excess of 8~\AA. All these 23 sources were included in our
primary sample of 51 sources selected according to the criteria of 
\S~\ref{sec:photo}. We therefore estimate the efficiency of our 
selection procedure to be 45 \%.

\subsection{Properties of the new Taurus members}

\begin{table*}
  \centering
   {\footnotesize

\begin{tabular}{ llccccccccc } \hline\hline
 Name&ST&Av&log$_{10}$(L/Lsun)&Teff&I&I-z&J-H&H-K&K
 \\ \hline
ZZ Tau IRS&M5.25&2.4&-1.4627&3100&15.95&0.85&1.40&1.13&10.31\\
KPNO-Tau 10&M5.50&0.0&-1.3214&3065&14.11&0.11&0.77&0.32&10.78\\
KPNO-Tau-2&M6.75&0.4&-2.1302&2889&16.56&0.49&0.68&0.49&12.75\\
KPNO-Tau-1&M9.00&0.4&-2.6340&2571&18.15&0.54&0.85&0.47&13.77\\
KPNO-Tau-6&M9.00&0.9&-2.5129&2571&17.90&0.66&0.80&0.51&13.69\\
KPNO-Tau-4&M9.50&2.5&-2.3744&2500&18.75&0.71&0.97&0.74& 13.28 \\ \\ \hline\\
CFHT-Tau 21&M1.25&6.6&-0.4185&3665&15.17&0.80&1.54&1.03&9.01\\
CFHT-Tau 19&M5.25&7.3&-1.1423&3100&16.54&1.15&1.79&1.52&10.54\\
CFHT-Tau 20&M5.50&3.6&-0.8586&3065&15.10&0.79&1.14&0.73&9.81\\
CFHT-Tau 17&M5.75&6.5&-1.1652&3030&17.62&0.82&1.58&0.88&10.76\\
CFHT-Tau 18&M6.00&4.9&-0.3715&2994&14.63&0.97&1.30&0.77&8.73\\
CFHT-Tau 10&M6.25&3.6&-1.6792&2959&16.85&0.99&1.09&0.60&12.13\\
CFHT-Tau 9&M6.25&0.9&-1.6218&2959&15.35&0.77&0.69&0.43&11.76\\
CFHT-Tau 7&M6.50&0.0&-1.2149&2924&14.12&\ldots&0.73&0.39&10.40\\
CFHT-Tau 8&M6.50&1.8&-1.6940&2924&16.43&\ldots&1.05&0.68&11.45\\
CFHT-Tau 12&M6.50&3.4&-1.4401&2924&16.26&1.03&1.01&0.59&11.55\\
CFHT-Tau 11&M6.75&0.0&-1.5964&2889&14.88&0.71&0.59&0.35&11.59\\
CFHT-Tau 13&M7.25&3.5&-2.1012&2818&17.90&1.05&0.86&0.52&13.45\\
CFHT-Tau 6&M7.25&0.4&-1.6223&2818&15.40&0.53&0.80&0.47&11.37\\
CFHT-Tau 5&M7.50&9.2&-1.1258&2783&18.79&1.09&1.74&0.94&11.28\\
CFHT-Tau 14&M7.75&0.6&-1.7494&2747&15.62&0.88&0.70&0.42&11.94\\
CFHT-Tau 15&M8.25&1.3&-2.4380&2677&17.94&1.08&0.69&0.55&13.69\\
CFHT-Tau 16&M8.50&1.5&-2.4152&2642&17.91&1.07&0.72&0.54&13.70\\
 \hline

\end{tabular}
 }

  \caption{Photometry and derived parameters for previously known (top) and new (bottom) Taurus \BD\ and VLM members}
  \label{tab:results}
\end{table*} 

\begin{figure}[t]
\includegraphics[width=\hsize]{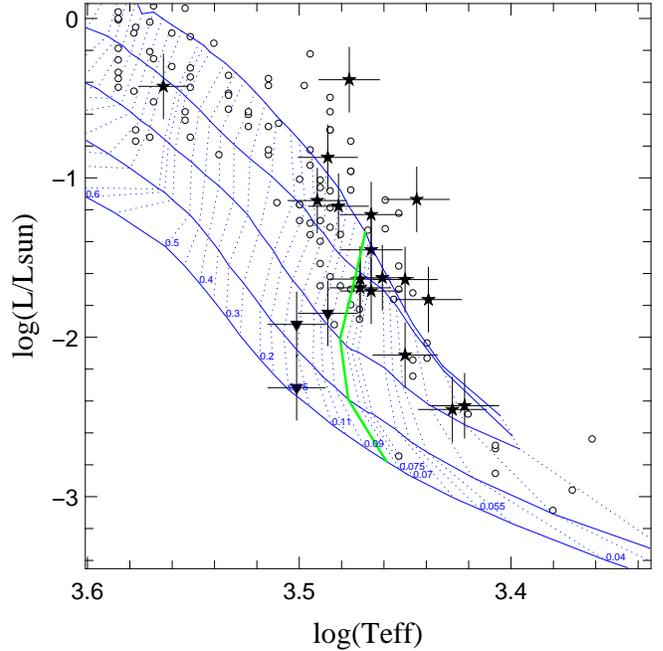}
\caption{Derived H-R diagram in Taurus. New Taurus members identified in this work
  are indicated with {\it Star} symbols. Previously known Taurus members
  are shown as {\it open circles}. We have also plotted with {\it
  triangles} the 3 sources with uncertain membership discussed in
  \S~4.1.  The pre-main sequence tracks from \cite{Chabrier-2000}
  (DUSTY models) are shown as blue curves: solid lines represent
  isochrones with ages of 1, 3, 10, 30 and 100 Myr, dotted lines
  represent mass tracks ranging from 0.02 M$_{\odot}$ to 1.4
  M$_{\odot}$ as labelled. The green curve shows the computed
  stellar/substellar boundary at 0.08 M$_{\odot}$.}
\label{fig:hr}
\end{figure}

We summarize in Table \ref{tab:results} the derived properties of the
Taurus members. We list the spectral types and reddening values derived
through our spectral fitting procedure. The effective temperature is
estimated from the spectral type using  the following linear fit to 
the relationship derived by \cite{Luhman-1999} for young M dwarfs: 
$T_{eff} = 3780 -175.6*SpT$. We
compute the luminosity from the dereddened I band magnitude, using the
bolometric corrections for M dwarfs from \cite{Luhman-1999}. We also list
our $I$ and $z^{\prime}$ band photometry combined with the J, H and K 2MASS
photometry. We show in Figure \ref{fig:hr} the derived HR diagram for our
new Taurus sources, including previously known members. We use the 0.08
M$_{\odot}$ mass track from the \cite{Chabrier-2000} pre-main sequence
models to define the stellar/substellar boundary, which falls at spectral
type M6.25V for a typical Taurus age of 3 Myr. We will therefore use in
the following this spectral type as the limiting boundary for brown dwarfs
in our sample. We note in the substellar domain a general trend of the
models to predict larger ages for the lowest mass objects. This tendency is
likely due to uncertainties in the pre-main sequence tracks and/or
bolometric corrections for the lowest masses.

We recover in our sample 6 previously known Taurus members, and we find 12
new \BDs\ and 5 new very low mass (\VLM) stars.  We show in Figure
\ref{fig:spec} the individual spectra of the 17 new Taurus members.
We also show ZZ~Tau~IRS, for which, to our knowledge, no low-resolution
optical spectrum is published in the literature. Four of the new \BDs\ were previously
published in conference proceedings by \cite{Monin-2004} and
\cite{Guieu-2004}. Two of these, CFHT-Tau~6 (J04390396+2544264) and
CFHT-Tau~8 (J04411078+2555116), have independently been discovered by
\cite{Luhman-2004}. This author found the same spectral type for CFHT-Tau~6
(M7.25) and a close extinction value ($\AV$=0.26). For CFHT-Tau~8  we
derive an M6.5 spectral type with a moderate reddening ($\AV$=1.8) while
\cite{Luhman-2004} found an M5.5 spectral type with A$_J$=0.7 ($\AV$=2.6).
We note that the discrepancy in $\AV$ values is similar to our estimated
reddening uncertainty (0.8 mag) while the difference of 1 spectral class is
twice our estimated spectral typing error.

Among the remaining Taurus members, our estimates of spectral type and
reddening values appear generally compatible, within the quoted uncertainties, with
values given in the literature, except for 2
sources for which larger discrepancies are found: ZZ~Tau~IRS and
KPNO~Tau~4. ZZ~Tau~IRS, the suspected energy source of HH~393
\citep{Gomez-1997}, has been recently observed spectroscopically by
\cite{White-2004}. These authors derive from their high-resolution optical
spectroscopic observations an M4.5$\pm$2 spectral type, compatible within
the quoted uncertainties with our own derivation of M5.25. However, using 
the (J-H) color, they estimate a reddening value of 7.6, significantly
larger than our own estimate of $\AV$ = 2.4, derived from the spectral
fitting procedure. We note that this source displays a significant excess
in the (J-H)/(H-K) diagram as well as strong accretion signatures, which
may significantly affect the derivation of reddening from near-infrared
photometry. Indeed, if we deredden ZZ~Tau~IRS back to the locus of
accreting T~Tauri stars in the $(J-H)/(H-K)$ diagram, we find $\AV = 3.1$,
a value compatible with our derivation from fitting of the optical low
resolution spectrum. We note however that these values may be a lower limit
to the true reddening if indeed this source is viewed edge-on as suggested
by its unusually large emission lines equivalent widths, as already noted
by \cite{White-2004}. We come back to this issue in the following section.
In the case of KPNO~Tau~4, we find the same spectral type (M9.5) but a
reddening value larger than the one computed by \cite{Briceno-2002} ($\AV$=
2.5 {\it vs} 0.0). This discrepancy likely arises from the 
rapid increase of the continuum flux longwards of $\simeq~8200$ \AA\
for spectral types later than M9V. This part of the continuum therefore 
strongly influences the derivation of reddening values at late spectral
types. Uncertainties on its exact shape result in large uncertainties on
derived $\AV$. We therefore estimate that for our later spectral types
($>M9$), uncertainties on derived $\AV$ values are almost doubled and
likely increase to 1.6 mag.

\subsection{Accretion and outflow signatures}
 
\begin{figure}[]
\includegraphics[width=\hsize]{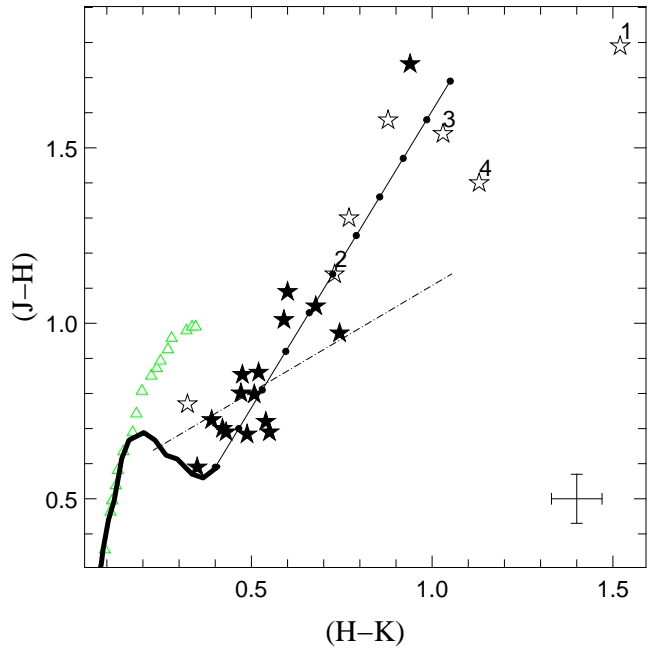}
\caption{(J-H)/(H-K) diagram. VLM stars (open star symbols) and brown dwarfs 
  (filled star symbols) Taurus members identified in this
  work are represented. Numbers identify the 4 VLM stars
  with intense line emission from Figure \ref{fig:line}. 
Also shown are the locations of the giant
  branch (open green triangles) and the dwarf sequence (thick black curve)
  extending from G2 to M7 (SpT). The black thin line illustrates a
  reddening vector of 10 $\AV$ extending from the M7 V dwarf colors. The
  locus of the accreting classical T~Tauri stars from \cite{Meyer-1997} is
  displayed with a dotted-dashed line.}
\label{fig:jhhk}
\end{figure}

\begin{figure}[]
  \includegraphics[width=\hsize]{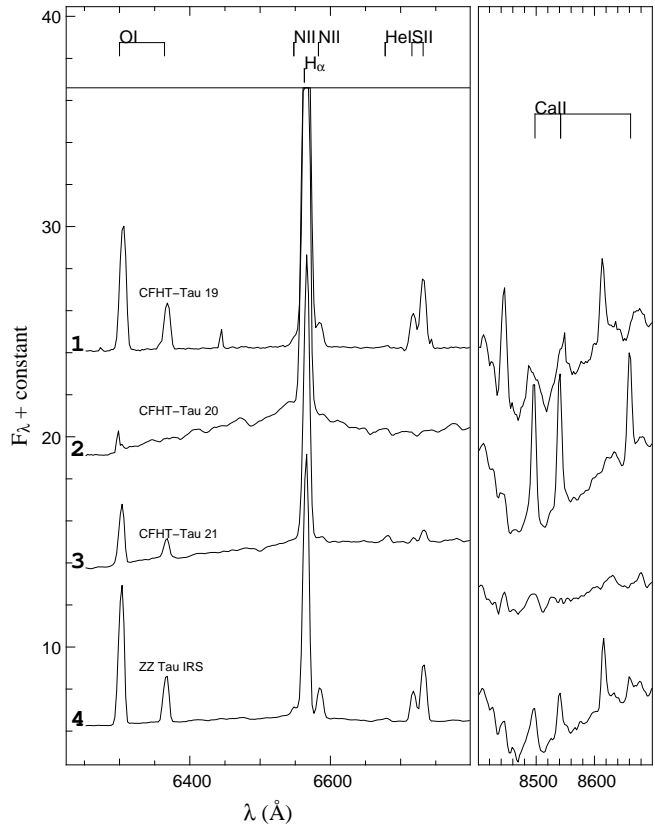}
  \caption{A zoom on representative regions of the spectra is presented for
    the 4 Taurus members with strong emission lines. }
  \label{fig:line}
\end{figure}

\begin{figure}[]
\includegraphics[width=\hsize]{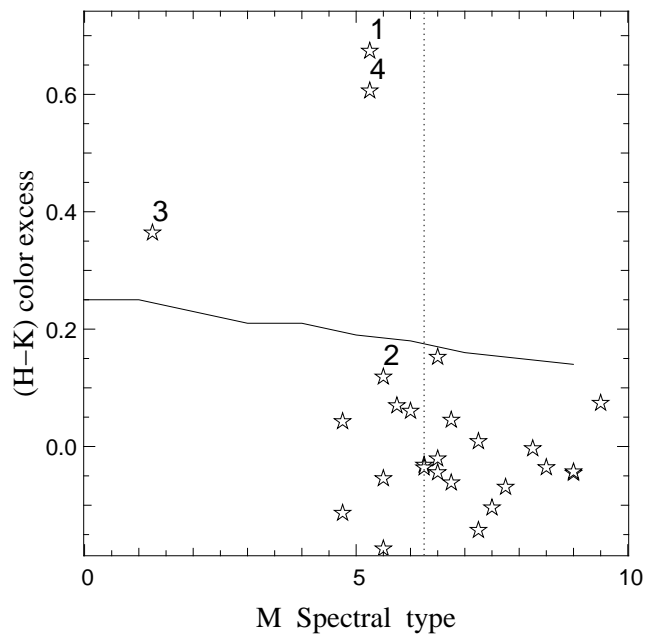}
\caption{(H-K) color excess versus M spectral type for
Taurus members from this study. VLM stars with intense line
emission are numbered as in Figure \ref{fig:line}. The solid line shows the
maximum (H-K) color excess expected from a flat reprocessing circumstellar disk
around a M0V to M9V central source \citep{Liu-2003}.}
\label{fig:excess}
\end{figure}

\begin{figure}[]
\includegraphics[width=\hsize]{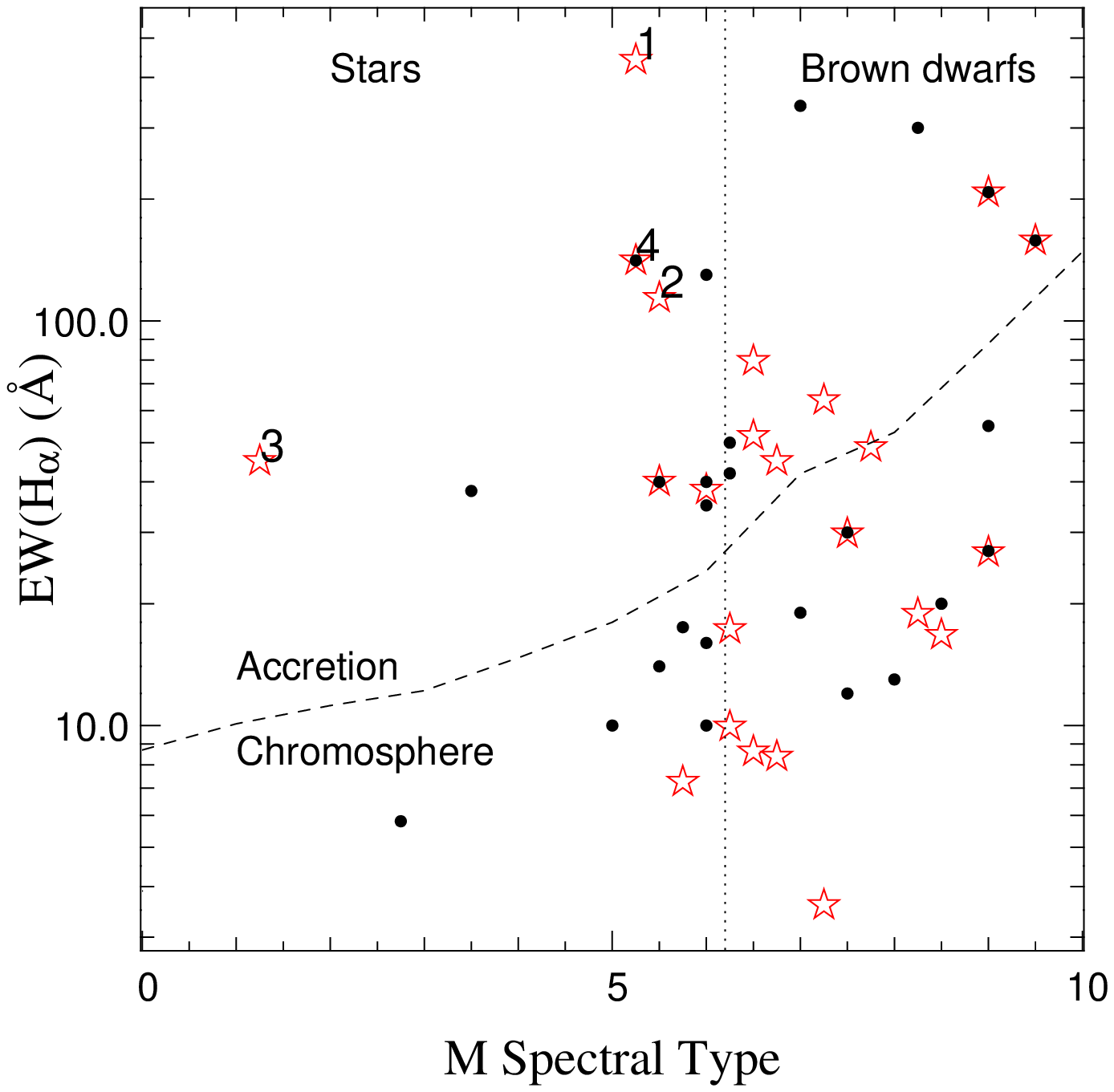}
\caption{$\Ha$ equivalent width versus spectral type for
  \VLM\ and \BD\ Taurus members. Star symbols identify the Taurus members
  studied in this paper. Previously known members are shown with black
  dots.  The dashed line shows the empirical boundary between chromospheric
  and accretion-related $\Ha$ activity derived by \cite{Barrado-2003}. The
  EW($\Ha$) is plotted on a logarithmic scale.}
\label{fig:ewha}
\end{figure}

  We examine the Taurus \VLM\ members identified in this work for
  evidence of accretion/outflow. These signatures include strong
  near-infrared excess, suggestive of an active accretion disk; strong
  emission in the optical forbidden lines which, in Classical T Tauri
  stars, trace outflow signatures strongly correlated with the
  accretion process \citep{Hartigan-1994,Cabrit-1990}; strong
  $\Ha$ emission beyond the limit for chromospheric activity. A large
  fraction of the newly identified Taurus \VLM\ members show
  indications of one or more of these accretion/outflow signatures.

   Three of the \VLM\ Taurus members (CFHT-Tau~19, ZZ~Tau~IRS,
  CFHT-Tau~21 ) and one of the \BDs\ (KPNO-Tau~4) exhibit near-infrared
  excess in the (J-H) versus (H-K) diagram (Figure~\ref{fig:jhhk}). These
  sources lie in this diagram to the right of the reddening vector
  extending from the M7 dwarf color and above the locus of accreting
  T~Tauri stars, suggesting near-infrared excess emission arising from the
  presence of an accretion disk. Using our estimates of reddening and
  spectral type, we further compute a $\Delta$(H-K) excess for each
  confirmed Taurus member. This near-infrared excess is represented in
  Figure \ref{fig:excess} as a function of spectral type and compared to
  the maximum (H-K) color excesses predicted for a flat reprocessing
  circumstellar disk around a central star of spectral type ranging from
   M0~V to M9~V \citep{Liu-2003}. The three \VLM\ members previously identified
  in the (J-H)/(H-K) diagram show a clear (H-K) color excess above what is
  expected from a pure reprocessing disk. Given the one $\sigma$ uncertainty
  of 0.1 mag in $\Delta$(H-K), we do not find evidence from this diagram
  for strong active accretion in the substellar population. However, the lack of
  detectable (H-K) excess does not imply an absence of disk
  around these sources. Indeed, detailed modelling of the spectral energy
  distributions from reprocessing disks around young substellar objects by
  \cite{Natta-2001} showed that excess disk emission becomes clearly
  detectable longward of 3 $\mu$m only.

The three sources with strong near-IR excess also display strong emission
in the optical forbidden lines of [O~{\sc i}]~$\lambda$6300, 6363~\AA, [N~{\sc ii}] 
$\lambda$6548, 6583~\AA, and [S~{\sc ii}]~$\lambda$6716, 6731~\AA~
(Figure~\ref{fig:line}), which, in higher mass CTTs, trace outflow
signatures strongly correlated with the accretion process
\citep{Hartigan-1994,Cabrit-1990}. One of these sources, ZZ~Tau
IRS, shows anomalously large equivalent widths in the optical forbidden
lines (EW([O~{\sc i}]~$\lambda$6300~\AA) = -190 \AA, EW([S~{\sc
ii}]~$\lambda$6731+6716~\AA) = -77 \AA). Typical equivalent widths in
T~Tauri stars are on the order of a few tenths of \AA. Such large values in
ZZ~Tau IRS could be explained if the source is seen close to edge-on and
the continuum is therefore suppressed relative to the line emission,
arising from extended jet emission. The suggestion that ZZ~Tau~IRS is close
to edge-on has already been made by \cite{White-2004} on the same argument.
 One   \VLM\ object ,  CFHT-Tau~20 (J04295950+2433078)  
has significant Ca~{\sc ii} line emission, an
indicator of accretion activity \citep{Muzerolle-1998}. Indeed,
\cite{Mohanty-2005} demonstrated recently that the Ca{\sc
ii}~$\lambda$8662~\AA\ line flux correlates very well with the disk mass
accretion rate down to the substellar regime.

 As noted before, all but 2 of the Taurus members identified in this study
display strong $\Ha$ emission with equivalent widths in excess of 8~\AA.
This level of $\Ha$ emission is usually taken in higher mass T Tauri stars
(TTs) as a strong accretion indicator. Chrosmospheric $\Ha$ emission
activity is however expected to increase strongly towards late M
types. \cite{Barrado-2003} conducted an extensive study of $\Ha$ emission
in a sample of young M stars. They derived an empirical CTTs/WTTs boundary,
extended to substellar analogs, in the $\Ha$ equivalent width versus
spectral type diagram. This boundary is defined by the saturation limit for
chromospheric activity at $\log {({\rm L}_{\Ha}/{\rm L}_{\rm bol})}= -3.3$, determined from
study of a sample of young open clusters. We plot in Figure~\ref{fig:ewha}
the distribution of late-type Taurus members with respect to this
boundary. We include members from this study and previously known late-type
sources with published $\Ha$ equivalent widths measurements from
\cite{Luhman-2000,Martin-2001,Briceno-2002,Luhman-2003a}.
Figure~\ref{fig:ewha} suggests that 6 Taurus brown dwarfs among our sample
( CFHT-Tau~6, CFHT-Tau~11, CFHT-Tau~12, CFHT-Tau~8, KPNO-Tau-4
and KPNO-Tau-6) have level of $\Ha$ emission in excess of chromospheric
activity and are likely CTTs substellar analogs. All but one
(CFHT-Tau~17) of the \VLM\ members from this study lie above the
CTTS/WTTs boundary. In particular, the 4 \VLM\ sources with strong emission
lines (labelled 1 to 4), discussed in the previous paragraph, clearly stand
out in this diagram as strongly accreting sources. We note
that high-resolution spectroscopic observations that fully resolve the $\Ha$
profiles are required to confirm that accretion is indeed present in these sources.

 The near-infrared excess emission and line emission properties of the
Taurus \VLM\ members analyzed here are strikingly similar to the
accretion/ejection signatures observed in the higher mass CTTs, indicating
that the same processes are occurring down to the substellar limit. The
apparent large fraction of accreting sources in our \VLM\ sample is
probably a result of our selection procedure since our sample of objects
with spectral types earlier than M4~V is biased towards large near-IR
excess sources. The fraction of accreting brown dwarfs, as deduced from the
properties of their $\Ha$ emission, is approximatively 40\% (6/15) in our
sample and 42 \% (10/24) when we include previously known substellar
members. We note that the substellar sample studied here does not suffer
from the same bias as the higher mass objects and represents 72 \% of all
currently known \BDs\ in Taurus. This derived substellar accretor
fraction in Taurus is statistically consistent with the one derived by
\cite{Mohanty-2005} for sources with spectral type $\ge$ M5 from a detailed
analysis of accretion signatures of 32 \% $\pm$ 10 \%.
The fraction of accreting sources among the substellar population is
similar to the one in the higher mass T~Tauri stars, suggesting that inner
disk lifetimes around young Taurus \BDs\ appear on the same order or larger
than around higher mass TTs.

\subsection{The stellar to brown dwarf ratio revisited \label{sec:rss}}

Following \cite{Briceno-2002, Luhman-2003a, Luhman-2004}, we quantify the
relative numbers of brown dwarfs and stars in Taurus.  The completeness in
mass of our spectroscopic survey is set on one hand by the completeness limits of
the 2MASS observations (J=15.25, H=14.4, K=13.75), which have been combined
with the optical data to select substellar Taurus candidates, on the other
hand by the $i^\prime$ band limit of 20 of our spectroscopic observations.  Using the pre-main
sequence tracks from \cite{Chabrier-2000} (DUSTY models), we compute in both
cases a corresponding mass limit of $30\,M_J$ and reddening limit of
$A_V<4$ for a maximum age of 10\,Myr at the Taurus distance (140\,pc). 

We derive below the reddening and mass limited sample
of all known Taurus members and update the current estimate of the brown
dwarf to star ratio by including the new fields studied in
this paper.

Compiling the list of known Taurus members from
\cite{Kenyon-1995,Briceno-1998,Briceno-1999,Briceno-2002,Luhman-1998,
Luhman-2000,Luhman-2003a,Luhman-2004,White-2004,Strom-1994,Martin-2000,Martin-2001,Wichmann-2000}
and including the 17 new members from this work, we have identified 64
Taurus members (including known companions) projected towards our new
fields but not included in the combined surveys of \cite{Briceno-1998,
Luhman-2000, Briceno-2002, Luhman-2004}.  Following \cite{Briceno-2002},
binary systems with separations $\leq 2^{\prime\prime}$ are considered as
individual objects. We are therefore left with 54
sources. \cite{White-2004} recently provided estimates of spectral types
and reddenings from high-resolution optical spectroscopy for a sample of
embedded young stars in Taurus. We adopt their derived effective
temperatures, luminosities and $\AV$ values for Haro~6-5B and 5 embedded
IRAS sources included in our sample (IRAS~04016+2610, IRAS~04260+2642,
IRAS~04264+2433, IRAS~04278+2253A, IRAS~04303+2240). We are left with 11
embedded IRAS sources for which we currently lack accurate estimates of
spectral types and reddenings. \cite{White-2004} derive, for the embedded
Class I sources in Taurus for which they have accurate spectral types, an
average visual extinction of 10.5 magnitudes with a standard deviation of
4.2 magnitudes. The 11 IRAS sources are therefore also likely to be very
embedded and not to contribute significantly in numbers to our reddening
limited sample.  As a consistency check, we derive for these sources a
rough estimate of $\AV$ from the (J-H)/(H-K) diagram following the method
described in \S \ref{sec:reddening}. Among the 11 IRAS sources, only one
source ends up with $\AV \leq 4$.  We further exclude FT~Tau, a strong
continuum source \citep{Herbig-1988}, and IRAS~04016+2610 and ZZ~Tau~IRS
for which \cite{White-2004} present strong indications that these are
edge-on systems. Among the remaining 40 sources, 31 sources have $\AV \leq
4$ and are therefore included in the reddening limited sample, 9 of these
are \BDs.

J04202555+2700355, J04213459+2701388, J04284263+2714039 and
J0442713+2512164, sources discovered by \cite{Luhman-2004}, are not
included in the substellar/stellar ratio computed by this author
because they correspond to an initial and incomplete 225 deg$^2$
survey.  However, since they are also present in our fields, they are
included into our present estimate of the substellar to stellar
ratio. In addition, the new Taurus members, CFHT-Tau~17,
CFHT-Tau~5, CFHT-Tau~7 and CFHT-Tau~11 discovered in this
study, fall inside the fields of
\cite{Luhman-2000,Briceno-2002,Luhman-2004} but are not reported by
these authors.  CFHT-Tau~17 and CFHT-Tau~5 are embedded
($\AV=6.5$ and $\AV=9.2$) but CFHT-Tau~7 and CFHT-Tau~11 have
little reddening and are therefore included in our ratio.  We estimate a
spectral type of M6.5 for CFHT-Tau~8 different from
\cite{Luhman-2004} (M5.5), so this source falls in our sample under the
hydrogen burning limit. We note that CFHT-Tau~7 lies at a projected
distance of only 20$^{\prime\prime}$ from V928~Tau~A/B and could be a
wide companion to this binary system. 

We combine the additional reddening limited sample derived from this
study with the one defined by \cite{Luhman-2004} and compute an
updated substellar to stellar ratio in Taurus. We first use the same
mass completeness limit of 20 $M_J$ as previous studies for
comparative purposes: \\
$\rss = N(0.08-0.02 M_{\odot})/N(0.08-10 M_{\odot}) = 31/127 = 0.24 \pm 0.05.$
Using a mass completeness limit of 30$M_J$ instead results in $\rss = 29/127 = 0.23\pm 0.05$.

As already mentioned in \cite{Luhman-2004}, the derived $\rss$ ratio in
Taurus is likely an upper limit for possible remaining incompleteness at
 low stellar masses. Indeed, the distribution of spectral types in Taurus
shows an apparent deficit at spectral types M2 to M4
\citep{Luhman-2003b}. Searches for substellar objects in Taurus are, as this
study, typically complete for spectral types later than M4, while
previous optical and X-ray large-scale surveys of the Taurus cloud are
likely complete down to $\simeq$ M2. 
The increase of Rss with increasing spatial coverage in Taurus could
be due to a larger incompleteness at these early to mid-M spectral
types in the distributed population than in the more studied high
stellar density regions. However, we show in the next section that
this is likely not the case: the distributed population does not show
an apparent larger deficit of M2-M4 stars (with respect to the earlier
spectral type population). Moreover, the incompleteness at spectral
types M2-M4 in Taurus has a limited effect on the computed Rss ratio:
increasing by a factor 2 the known number of stars in this spectral
range would bring the Rss ratio down to 0.20 +/- 0.05, compatible
within one sigma with our new estimate.

Our new estimate of the Taurus substellar to stellar ratio of 0.23 +/- 0.05
is higher than the value derived by Luhman (2004) of 0.18 +/- 0.04, although
still compatible at the 1 sigma level. It is now however comparable to the
Trapezium value of $\rss(\mbox{Trapezium}) = 0.26\pm 0.04$ 
estimated by \cite{Briceno-2002}, using the same evolutionary models
and treating binary systems in the same manner. Recent estimates 
seem to suggest an even lower value for the Trapezium. 
\cite{Slesnick-2004} have recently computed a substellar mass function for
the Orion nebula cluster with a sample of spectroscopically confirmed
members. These authors, using  a different pre-main sequence model
\citep{d'Antona-1997}, found a ratio of $\rss$=0.20, lower than the previous estimates
derived from photometric studies alone \citep{Luhman-2000,Hillenbrand-2000,Muench-2002}.

\subsection{The spatial distribution of \BDs}

\begin{figure*}[b]
  \includegraphics[width=\hsize]{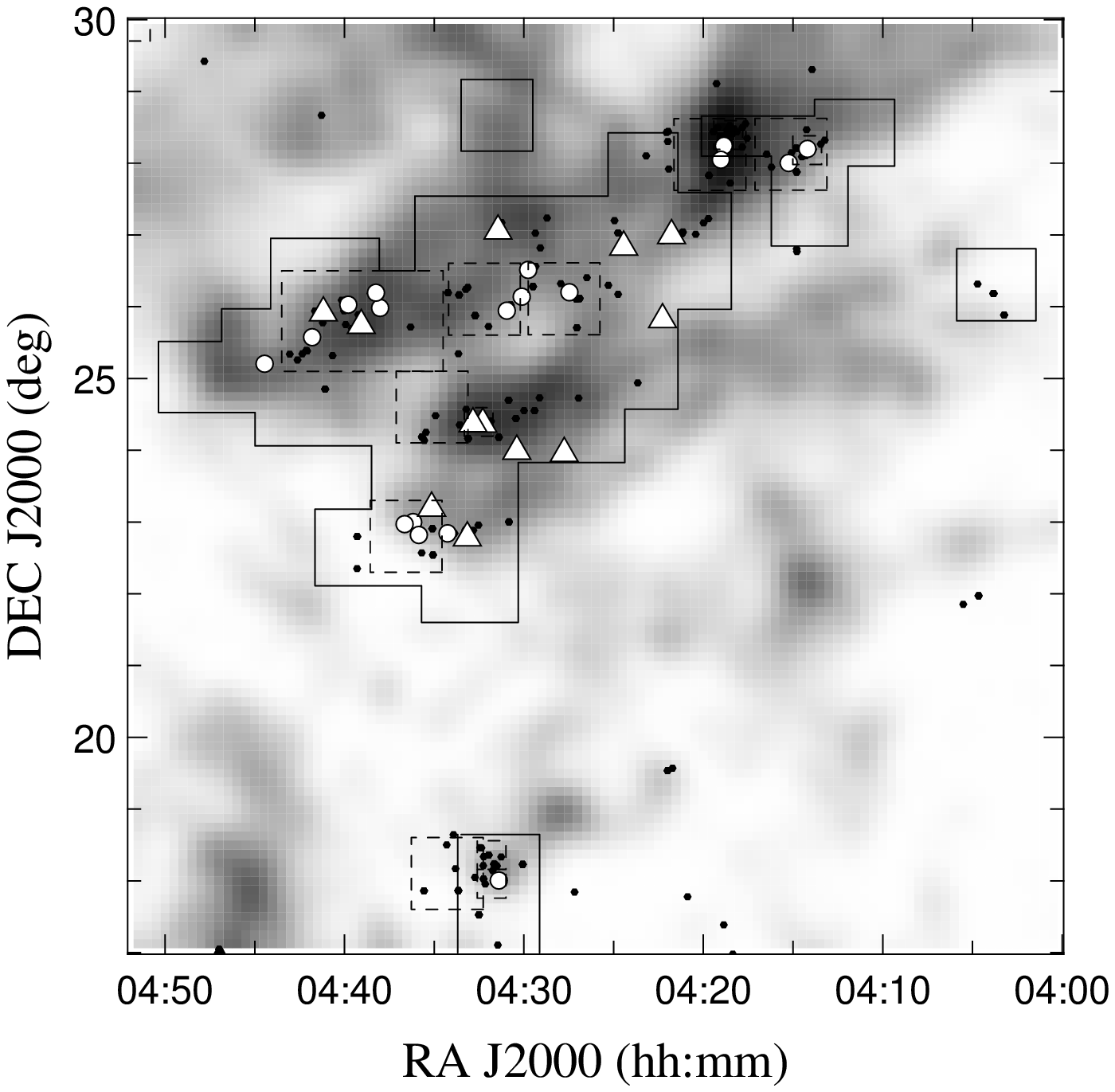}
  \caption{The spatial distribution of the newly found Taurus \BDs\
(sources with SpT $\le$ M6.25, triangle symbols) is shown superposed upon a
Taurus CO map from \citet{Ungerechts-1987} (grey scale). Black dots
indicate the distribution of previously known Taurus stars; filled white
circles correspond to previously known \BDs.  The black curve shows the
outline of our optical CFHT survey, while the spatial coverage of the previous surveys from \cite{Luhman-2000},
\cite{Briceno-2002} and \cite{Luhman-2004} is shown in dashed lines.}
\label{fig:distrib}
\end{figure*}

\begin{figure}
  \includegraphics[height=\hsize,angle=270]{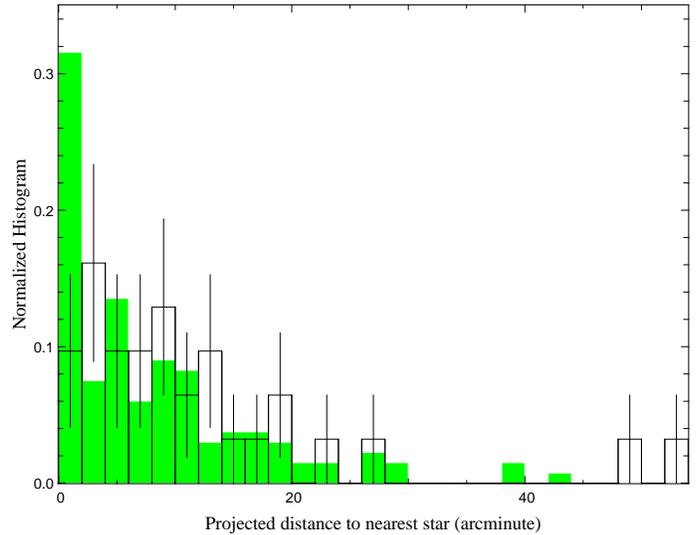}
  \caption{Normalized histograms of the distance to the nearest {\it
    stellar} neighbor, for stellar members only (filled histogram) and for the
    substellar population only (empty histogram).}
\label{fig:histdist}
\end{figure}

\begin{figure}
  \includegraphics[height=\hsize,angle=270]{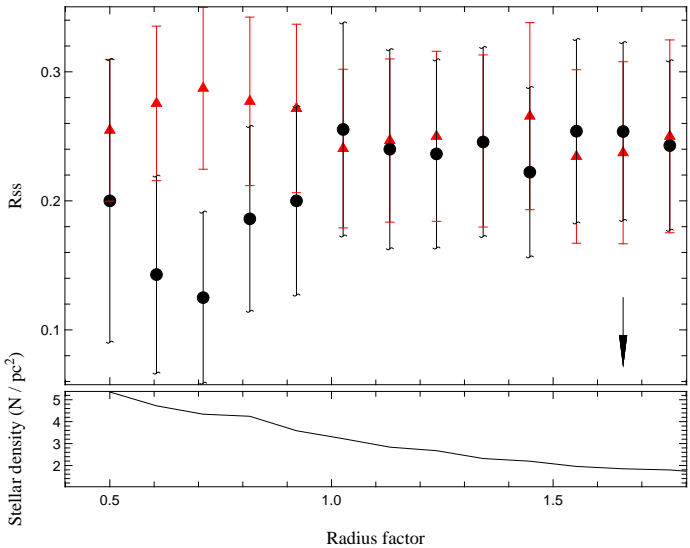}
  \caption{Variation of the ratio of substellar to stellar members in
the aggregates (black dots) versus the distributed population (red
triangles) as a function of the radius taken for the aggregates (in units
of the physical radius R$_c$ from \cite{Gomez-1993}). The vertical arrow
 corresponds to an equal number of members in the aggregates and
distributed population. The curve below shows the corresponding average
stellar density per pc$^{-2}$ within the aggregates.}
\label{fig:rss}
\end{figure}

Figure~\ref{fig:distrib} shows the spatial distribution of the currently
identified stellar and substellar members in Taurus, including all the new
sources identified in this work. Figure~\ref{fig:distrib} suggests that the
Taurus substellar members closely follow the distribution of stars. In
order to quantify this effect, we compute the distribution of the nearest
{\it stellar} neighbor, both for the stellar sample (defined as members
with spectral types earlier than M6.25V) and the substellar sample. In this
calculation, we consider only sources projected towards fields searched for
substellar members and all binary systems with separation less than
10$^{\prime\prime}$ (ie including virtually all known binary systems in
Taurus) as a single stellar source. This results in 184 stellar systems and
33 substellar members. The histogram of {\sl nearest stellar neighbor}
distances is shown in Figure~~\ref{fig:histdist} for both populations. The
two distributions appear statistically consistent. The median of the
distribution for the stellar sample is: 4.82$^{\prime}$ with a standard
deviation of 8.45$^{\prime}$, while for the substellar sample we find a
median of 5.72$^{\prime}$ (standard deviation
10.48$^{\prime}$). Considering only the sources with $\AV\le 4$ does not
significantly change the result. 

However, the fact that the substellar to stellar ratio ($\rss$) has been
steadily increasing as the area covered in Taurus became larger suggests a
dependency of this ratio on the stellar density. To quantify this effect,
we divide the Taurus population in two components: aggregates and
distributed population. To define the former population, we use the center
of the aggregates from \cite{Gomez-1993} and a fixed factor of the physical
radii R$_c$ determined for each aggregate by these authors, which range
from 0.5 to 1.1~pc. We plot in Figure~\ref{fig:rss} the variation of the
substellar to stellar ratio in both populations as the surface comprising
the aggregates is increased. In the computation of the $\rss$ ratio we
consider only sources with $\AV \le 4$ and binary systems with separations
$\ge$ 2$^{\prime\prime}$ are treated as a single source.
Figure~\ref{fig:rss} clearly shows a marked increase of the $\rss$ ratio as
the integration radii defining the aggregates are increased. For
integration radii $\simeq$ 0.7R$_c$, the $\rss$ ratio is a factor 2.3 lower
in the aggregates than in the remaining distributed population. At
integration radii $\ge$ R$_c$ both $\rss$ values converge towards the
average value of 0.24 determined earlier.  Figure~\ref{fig:rss}
illustrates how $\rss$ increases from
$\simeq$ 0.13 \citep{Briceno-2002}
towards the higher stellar density aggregates, to 0.24 (this work)
as larger and lower stellar surface density areas are being
surveyed. Figure~\ref{fig:rss} suggests a \BD\ deficit (of a factor
$\simeq$ 2) in the central regions of the Taurus aggregates, on spatial
scales $\simeq$ 0.35 to 0.7 pc, where stellar densities are the highest. 

As mentioned above, this apparent increase of $\rss$ could be due to a larger
incompleteness in spectral types M2V to M4V in the distributed population
than in the more studied aggregates. To test this hypothesis, we have
computed the evolution of the ratio $\frac{N(M2 < Sp < M4)}{N( SpT <
M2)}$. We find that this ratio is lower 
in the cores of the aggregates than in the distributed population and
increases as the average stellar density decreases. 
We conclude that there is no strong reason to believe that the incompleteness 
increases away from the aggregates and  
that the observed increase of the $\rss$ ratio with decreasing
average stellar density in Taurus is real.

\section{Discussion: implications for substellar formation}

 Two main classes of models have been proposed for the formation of
substellar objects. In the {\sl standard formation scenario} brown dwarfs
form like stars, through (turbulent) gravitational collapse and
fragmentation of cores, followed by subsequent disk accretion. In
the {\sl ejection models}, brown dwarfs are stellar embryos ejected from
their parent core either early in their evolution from dynamically unstable
multiple protostellar systems \citep{Reipurth-2001} or through secular dynamical
decay in dense embedded clusters \citep{Sterzik-2003, Kroupa-2003a}. Thus
deprived of a significant mass reservoir, these ejected embryos end up as
very low mass objects. We discuss below the implications of our
observational results on the substellar population of the Taurus cloud for
these two main classes of models.

  The fact that the abundance of \BDs\ (down to 30 M$_{\rm J}$) relative to
stars is found to be the same ($\simeq$ 25 \%) in the diffuse Taurus cloud
and the high-density ONC cluster seems to suggest that there is no strong
dependency of the substellar IMF on initial molecular cloud conditions, in
particular gas density and level of turbulence. This result seems to be in
contrast with the most recent simulations of supersonic turbulent
fragmentation by \cite{Padoan-2004,Delgado-2004}, with sonic rms Mach
numbers in the range 3 to 20, which predict a strong dependency of the
abundance of \BDs\ with the initial cloud velocity dispersion and gas
density. In the subsonic case, \cite{Goodwin-2004a} show that 20 \% of the
objects formed are brown dwarfs almost independently of the level of
turbulence. It is unlikely however that a subsonic model would apply to the
ONC precursor where supersonic turbulent motions were likely present.
It should be noted that turbulent fragmentation models predict a decrease
of the average mass formed with an increasing level of turbulence, which
could account for the difference in peaks in the high mass end of the mass
distributions of the Taurus and ONC populations.

 Although we do not find statistical evidence for a spatial segregation
between stars and brown dwarfs in Taurus, we do find strong indication that
the abundance of \BDs\ relative to stars in the centers of the Taurus
aggregates, on scales $\simeq$ 0.5~pc, is lowered by a factor $\simeq$ 2
with respect to the more distributed population. An opposite trend would be
expected in fragmentation models where the Jeans mass decreases with
increasing gas density, leading to a corresponding increase in the
abundance of \BDs. This result seems to be best explained if a fraction of
the distributed population in Taurus is formed of low-mass stars and
substellar objects ejected from the aggregates, either through dynamical
encounters in a dense cluster ({\sl the collision} model) or through rapid
dynamical decay in unstable small N-body systems ({\sl the embryo-ejected}
model).  

\cite{Kroupa-2003a} study the secular dynamical evolution of Taurus-Auriga
type aggregates. They show that expansion occurs on timescales of a few
Myr, corresponding to a few crossing times, due to two body encounters and
gas removal. However, at the median age of the Taurus population ($\simeq$
3 Myr) the \BD\ population does not separate significantly from the stellar
population in these simulations. Long timescales are required, $t \ge 10
Myr$, for a significant gradient in the relative number of \BDs\ to
develop within a central 1pc sphere.  Thus, secular dynamical evolution
seems to require too long timescales to account for the present spatial
distributions of stars and BDs in Taurus.
\cite{Kroupa-2003b} have also investigated the case where \BD\ ejections
originate from dynamically unstable small N-systems (the {\sl
embryo-ejection} hypothesis). The decay of small N groups occurs on time
scale of a few 10$^{4}$ yr, ie the production of most \BDs\ occur well
before the end of gas removal, and the resulting escape velocities are
larger than in the previous scenario. The authors find that the number of \BDs\
per star seen in the Taurus survey of \cite{Briceno-2002}, taking into
account \BDs\ that escaped detection from their fields, and the central
region of the ONC cluster are best explained if in both cases {\sl the
embryo ejection model} dominates and produces the same abundance of ejected
\BD\ per star of 25 \% with a dispersion velocity $\le $ 2 km s$^{-1}$. The
authors predict than in Taurus-Auriga, on scales $\simeq$ 1 pc from the
center of the aggregates, there should be about twice as many unbound
\BDs\ than \BDs\ bound to the aggregates. This scenario appears in good
agreement with our observational results.

Recent numerical simulations also seem to support this scenario.
\cite{Goodwin-2004b} follow the fragmentation of a distribution of 5
$\Msun$ cores with turbulence properties similar to the Taurus cores. The
resulting mass function is bimodal. Roughly 50 \% of the sources,
predominantly the low-mass ones, are ejected from the cores through rapid
dynamical evolution before they could accrete significantly, and form a
flat low-mass distribution component. A high-mass log-normal component
results from objects which remain bound in the core and grow by
accretion. The ejected population comprises twice as many low-mass stars as
substellar objects. Ejection velocities are 1-2 km s$^{-1}$ almost
independent of mass, leading to the same spatial distribution of \BDs\ and
low-mass stars. As mentioned before, in these simulations the abundance of
\BDs\ relative to stars is on the order of 20 \%, independent of the exact
level of turbulence, and in close agreement with the Taurus observations.
 
  The line emission and infrared excess emission properties of the very low
mass population in Taurus indicate that accretion/ejection processes
similar to the ones observed in the higher mass T~Tauri stars proceed
down to the substellar regime and that the inner disk lifetimes is
similar in brown dwarfs and stars. These results have been taken by
different authors as indication that brown dwarfs form through the same
processes as stars and as an argument against ejection models.  However,
the survival of the inner disk in young brown dwarfs may not be
incompatible with the ejection model. Indeed with the typical inferred
accretion rates of $10^{-10}$- $10^{-11} \Msun {\rm yr}^{-1}$, and with a
typical disk mass of 1 \% of the central stellar mass, survival times for
an accretion disk truncated at R$_{\rm out}$ = 10~AU around a 50 M$_{\rm J}$ \BD\
are a few Myr \citep{Liu-2003}, ie on the same order as the age of the
Taurus population. Viscous evolutionary time-scales appear to be on the
same oder.  Indeed, $t_{\nu} \simeq \frac{1}{\alpha \Omega} (\frac{R}{H})^2$
(e.g. \cite{Armitage-1997}) scales as $M^{-1/2}$ where $M$ is the central
stellar mass. Viscous time-scales for disks of R$_{out}$ = 10~AU around
brown dwarfs of 50 M$_J$ are on the order of $ 2 \times 10^{6}$ yr (for
$\alpha$ values of 10$^{-3}$ at r=10 AU). Thus accretion signatures may
well last for a few Myr around brown dwarfs, even if small disks, with
outer radii on the order of 10~AU, surround them, as expected if they
have been ejected early in their evolution from their birth site \citep{Bate-2002}.

In summary, the {\sl embryo-ejection} model reproduces both qualitatively
and quantitatively the spatial distribution of stars and BDs in Taurus as
well as the evolution of the $\rss$ ratio and does not appear incompatible
with the accretion properties of the Taurus substellar population.

\section{Summary and conclusions}

We have presented in this work results from a spectroscopic follow-up study
of 79 very low mass Taurus candidates, selected from a large scale optical
(I,z$^{\prime}$) survey covering a total area of $\simeq$ 30 square
degrees. Our spectroscopic survey is 90 \% complete for photometric
candidates later than spectral type M4V, 100 \% complete for substellar
candidates (spectral types later than M6.25V). Our corresponding mass
completeness limit is 30 M$_J$ for ages $\le$ 10 Myr and $\AV\ \le 4$.  We
derive reddening values, spectral types and luminosity classes from a
spectral fitting procedure and Na {\sc i} equivalent widths measurements.
We identify 17 new Taurus members among which 12 are brown dwarfs. We
investigate their accretion properties, spatial distribution and abundance
relative to stars. The main results are the following:

\begin{enumerate}

\item A large fraction of the newly identified Taurus members show
indication of accretion/outflow signatures. Two of the new \VLM\ members
(CFHT-Tau~19 and CFHT-Tau~21) display near-infrared excess emission
and optical forbidden lines indicative of accretion/ejection processes
similar to the ones observed in the higher mass classical T~Tauri stars. In
addition, 4 of the new substellar members (CFHT-Tau~6, CFHT-Tau~8,
CFHT-Tau~11 and CFHT-Tau~12) exhibit $\Ha$ emission in excess of
chromospheric levels and are likely CTTs substellar analogs. From levels of
$\Ha$ emission, we estimate a fraction of accreting sources of 42 \% 
in the Taurus substellar population, similar to what is observed in the
higher mass T~Tauri stars.  
 
\item We derive a new estimate of the substellar to stellar ratio in Taurus
of $\rss = \frac{N(0.03-0.08 \Msun)}{N(0.08 - 10 \Msun)} = 0.23 \pm 0.05$,
1.3 times higher than the last estimate from \citet{Luhman-2004} derived
from spatially less extended studies. This value now appears fully
consistent with the value derived with similar assumptions for the
Trapezium cluster population of $\rss = 0.26 \pm 0.04$ by
\cite{Briceno-2002} and suggests an average production rate of one \BD\ per
4 stars in both regions.

\item 
We find a strong indication that the abundance of \BDs\ relative
to stars in the centers of the Taurus aggregates, on scales $\simeq$
0.5~pc, is lowered by a factor $\simeq$ 2 with respect to the more
distributed population.

\end{enumerate}

The similarity of the abundance of \BDs\ with respect to stars in both the
Taurus and ONC populations seem to favor ejection models for the origin of
the substellar population. The similar spatial distributions of the Taurus
stellar and substellar populations but the detectable gradient in the \BD\
abundance relative to stars at a characteristic age of $\simeq$ 3 Myr favor
\BD\ ejection through dynamical decay of unstable small N systems (the {\sl
embryo-ejection model}). Further tests are however required to confirm this
hypothesis, in particular multiplicity studies of the distributed
population. In addition, our analysis relies solely on the relative
abundance of \BDs\ with respect to stars. Detailed and complete studies of
the shape of both the substellar and stellar IMFs in star-forming regions
of different environments are clearly required. 

\begin{acknowledgements}
We thank Estelle Moraux and J\'er\^ome Bouvier for enlightening discussions
about the origin of substellar objects as well as the ESO and Keck
observatory staff for their help during the observations. This research has
made use of the 2MASS and CDS database. This work is based upon research
supported by the National Science Foundation under grant AST 02-05862. Any
opinions, findings, and conclusions or recommendations expressed in this
material are those of the authors and do not necessarily reflect the views
of the National Science Foundation.  The authors wish to extend special
thanks to those of Hawaiian ancestry on whose sacred mountain of Mauna Kea
we are privileged to be guests. Without their generous hospitality, the
Keck~I telescope observations presented therein would not have been
possible.
\end{acknowledgements}
\bibliographystyle{aa}
\bibliography{mybiblio}

\end{document}